\documentclass[11pt,a4paper]{article}
\usepackage[utf8]{inputenc}
\usepackage{amsmath,amssymb}
\usepackage[english]{babel}
\usepackage{graphicx}
\usepackage{multirow}
\usepackage{wrapfig}
\usepackage{url}
\usepackage{graphicx}
\usepackage{tabularx}
\usepackage[font={small,it}]{caption}
\usepackage{listings}
\usepackage{scalerel}
\usepackage{calligra}
\DeclareMathAlphabet{\mathcalligra}{T1}{calligra}{m}{n}
\usepackage{float}
\usepackage{amsfonts}
\usepackage{xcolor}
\usepackage{physics}
\usepackage[normalem]{ulem}
\usepackage{hyperref}
\usepackage{mathtools}
\usepackage{tcolorbox}
\usepackage{soul,xcolor}
\usepackage[toc,page]{appendix}
\usepackage{a4wide}
\usepackage{csquotes}

\usepackage{pdflscape}
\usepackage{parskip}
\usepackage{geometry}
\usepackage{booktabs}
\usepackage{tabularx}
\usepackage{makecell}
\definecolor{rowblue1}{RGB}{238,246,255}
\definecolor{rowblue2}{RGB}{218,236,255}
\definecolor{rowblue3}{RGB}{190,219,247}
\definecolor{rowblue4}{RGB}{145,190,230}
\usepackage{array}
\newenvironment{aleq}
    {\begin{equation}\begin{aligned}}
    {\end{aligned}\end{equation}\ignorespacesafterend}
\usepackage{subfiles}
\graphicspath{{figures/}}
\usepackage{subcaption}
  \usepackage{a4wide}
  \usepackage{latexsym}
  \usepackage{epsf}
  \usepackage{amssymb}
  \usepackage{graphicx}
  \usepackage{amsmath, cite}
  \usepackage{amsmath,amssymb,amsthm}
  \usepackage{verbatim}
  \usepackage{hyperref}
  \usepackage{color}
  \usepackage{mathtools}
  \usepackage{soul,xcolor}

\usepackage[table]{xcolor}
\usepackage{array}
\usepackage{booktabs}
\usepackage{tabularx}
\usepackage{graphicx}

\definecolor{rowblue1}{RGB}{238,246,255}
\definecolor{rowblue2}{RGB}{218,236,255}
\definecolor{rowblue3}{RGB}{196,224,255}
\definecolor{rowpurple}{RGB}{218,210,255}

\newcolumntype{C}{>{\centering\arraybackslash}X}

\usepackage{makecell}

\newcommand{\be}{\begin{equation}}
\newcommand{\ee}{\end{equation}}

\begin{document}
\numberwithin{equation}{section}

\vspace{2.718cm}
\begin{center}

{\LARGE \bf{
Low-energy brane decoupling in AdS flux vacua}}\\

\vspace{1cm}

 {\large Fien Apers $^a$ and Noelia Sánchez González $^b$}\\
 \vspace{0.5 cm}
 {\small  $^a$Instituto de F\'{i}sica Te\'{o}rica IFT-UAM/CSIC,
C/ Nicol\'{a}s Cabrera 13-15, Campus de Cantoblanco, 28049 Madrid, Spain}\\
\vspace{0.2 cm} {\small\slshape $^b$Rudolf Peierls Centre for Theoretical Physics
Beecroft Building, \\Clarendon Laboratory, Parks Road, University of Oxford, OX1 3PU, UK}\\
\vspace{0.5 cm} {\small\slshape fien.apers@uam.es, noelia.sanchezgonzalez@physics.ox.ac.uk}\\

\vspace{1cm}

 \end{center} {
}

\begin{abstract}
We study the decoupling of branes sourcing AdS flux vacua from the asymptotic bulk
using scalar-wave absorption probabilities. In a one-modulus truncation, the result
depends only on the dimension of the AdS vacuum and on the asymptotic steepness of
the scalar potential. We find that \textcolor{black}{for any steepness and for any dimension greater than two} the absorption probability vanishes in the
low-energy limit, showing that the branes decouple from the bulk.

We illustrate the general analysis in several scale-separated AdS vacua. For DGKT
we find $\mathcal{P}_{\rm abs}\sim\omega^{27/7}$ and an effective transverse
dimension $d_{\rm eff}=20/7$. Interestingly, the scalar potential along the
brane-induced moduli trajectory agrees with that obtained by compactifying a
hypothetical $(4+13/7)$-dimensional gravity theory on a $13/7$-dimensional sphere
threaded by flux. We also apply the analysis to scale-separated AdS$_3$ examples
and to the simplest one-modulus KKLT model.
\end{abstract}
\newpage

\tableofcontents
\newpage
\section{Introduction}
Scale-separated AdS vacua are anti-de Sitter (AdS) vacua in which the Kaluza--Klein (KK) scale associated with the extra compact dimensions is parametrically separated from the AdS curvature scale. Such a hierarchy is required for a genuine lower-dimensional effective description and is therefore arguably one of the most basic phenomenological conditions one may impose on a string compactification. Nevertheless, whether parametrically scale-separated AdS vacua can actually arise in string theory remains an open and actively debated question.

One source of skepticism comes from the Swampland program, which has emphasized the subtleties of taking asymptotic limits in field space. Parametric scale separation typically requires approaching precisely such an extreme limit. This observation has motivated several conjectures related to the Distance Conjecture \cite{Ooguri:2006in}, including the AdS Distance Conjecture and the Strong AdS Distance Conjecture \cite{Lust:2019zwm}, as well as the Refined Strong AdS Distance Conjecture \cite{Buratti:2020kda}, which constrain how the KK scale can depend on the AdS Hubble scale.

At the same time, several candidate constructions of scale-separated AdS vacua are known. Among the best-studied classical examples are the DGKT-CFI vacua \cite{DeWolfe:2005uu,Camara:2005dc}, together with more recent AdS$_3$ constructions that share many of their features \cite{Farakos:2020phe,Arboleya:2024vnp,Farakos:2025vkn,VanHemelryck:2025qok}. Beyond the classical regime, prominent candidates include KKLT \cite{Kachru:2003aw} and the Large Volume Scenario (LVS) \cite{Balasubramanian:2005zx}, where quantum effects play an essential role. None of these constructions is yet fully explicit in all respects, although substantial progress toward establishing their consistency has been made in recent years, \cite{Junghans:2020acz, Demirtas:2019sip,Demirtas:2021nlu, Emelin:2022cac}.

The question becomes particularly intriguing from the perspective of holography. In a putative dual conformal field theory, parametric scale separation in the bulk should manifest itself as a parametrically large gap in the spectrum of single-trace primary operators above the operators corresponding to the light bulk fields. No explicit \textcolor{black}{conformal field theory (CFT)} exhibiting the required properties is currently known \cite{Collins:2022nux}. Moreover, the candidate dual theories are expected to have unusually large central charges. In the DGKT construction, for example, the central charge scales as \cite{Aharony:2008wz}
\begin{aleq}
c \sim N^{9/2},
\end{aleq}
whereas in KKLT-type constructions the corresponding scaling can be exponential in the relevant control parameter.

Identifying the CFT duals of these scale-separated AdS vacua \cite{Aharony:2008wz,Conlon:2021cjk,Apers:2022tfm,Apers:2022zjx,Apers:2022zjx2} would therefore be valuable for two complementary reasons. If such duals exist, they should provide examples of CFTs with striking and potentially novel spectral properties. Conversely, constraints from conformal field theory and holography \cite{Alday:2019qrf,Perlmutter:2024yhb,Bobev:2023bxl,Bobev:2025,Revello:2026orbifold,Revello:2026dgkt,Revello:2026rigid,Lust:2022lfc,Bena:2024,Conlon:2018vov,Conlon:2020wmc} may offer a route toward proving that the proposed parametric limits cannot exist in a consistent theory of quantum gravity.

In this paper, we will focus on the brane duals of these scale-separated AdS vacua. These are stacks of D-branes probing a certain geometry whose near-horizon limit reproduces the AdS vacuum, with the putative dual CFT living on the worldvolume of these branes. There has been recent progress towards constructing such brane duals for AdS flux vacua in general \cite{Apers:2025pon, Apers:2026lgi}. At the same time, concerns have been raised about whether these brane theories can successfully decouple from the bulk, as required for a holographic description \cite{Bedroya:2025ltj,Bedroya:2025constraints}.

The concern raised by Bedroya and Steinhardt \cite{Bedroya:2025ltj,Bedroya:2025constraints} is that, for the brane description of a scale-separated AdS vacuum, perturbations originating from the branes become increasingly blueshifted as they propagate towards the asymptotic region. At sufficiently large radial distances, their energy can exceed the string scale, signaling a breakdown of the perturbative description. However, a recent paper by one of the authors of this work \cite{Apers:2026lgi} has offered a different perspective on this issue in the context of the DGKT brane dual, based on two observations. First, despite the presence of this blueshift in the asymptotic region, there is an infinite redshift between the DGKT branes and any observer at fixed radial distance. Consequently, perturbations originating from the branes appear infinitely redshifted, rather than blueshifted, to such an observer. Second, decoupling in standard examples of AdS/CFT is established by taking an appropriate low-energy limit \cite{Klebanov:1997kc, Maldacena:1997re}. An analysis of this limit for the DGKT brane dual shows that perturbations encounter an infinite potential barrier separating the branes from the asymptotic bulk, such that their transmission into the bulk is exponentially suppressed, as in standard examples. From this perspective, the low-energy dynamics of the DGKT brane dual supports a successful decoupling of the branes from the bulk.

In this paper, we extend the decoupling analysis of \cite{Apers:2026lgi}. We perform a detailed computation of the absorption probability and cross-section for scalar-wave perturbations originating from the brane. This provides more information than the presence of a potential barrier alone, as it allows us to determine the precise rate of decoupling. Moreover, the presence of a potential barrier in a particular coordinate system is only a sufficient condition for decoupling, rather than a necessary one, whereas the vanishing of the absorption probability in the low-energy limit provides a necessary criterion. In this way, we extend the decoupling analysis to AdS flux branes more generally and show that these branes generically decouple from the bulk in the low-energy limit. In particular, we obtain a general expression for the absorption probability and cross-section in terms of only two pieces of data: the dimensionality of the AdS spacetime and the steepness $\lambda$ of the asymptotic potential (for the flux branes considered here, the contributions to the potential arise from fluxes and are therefore exponential in the canonically normalised moduli; \textcolor{black}{we focus on domain wall ansatzes with one radial coordinate, naturally singling out one-modulus asymptotic limits.})

\textcolor{black}{In defining the absorption-cross-section, we encounter an interesting side result. 
The relevant phase-space factor is controlled by the effective dimensionality seen by the propagating waves. 
In familiar examples such as $\mathrm{AdS}_5\times S^5$, this has a direct geometric interpretation as the dimension of the transverse space. 
For the non-flat asymptotics of AdS flux vacua, however, no such interpretation is immediate, so we instead define an effective dimension directly from the asymptotic wave equation.}

\textcolor{black}{For generic flux vacua, this effective dimensionality can simply be viewed as a convenient way of characterizing the asymptotic scattering problem. 
Interestingly, for DGKT and related vacua with integer conformal dimensions 
\cite{Conlon:2021cjk,Apers:2022zjx,Apers:2022tfm,Quirant:2022fpn,Plauschinn:2022ztd,Apers:2022zjx2,Andriot:2023fss,Farakos:2025vkn}, 
it appears to reflect a deeper structural similarity with ordinary Freund--Rubin compactifications, potentially involving an effective fractional number of extra dimensions. 
We return to this interpretation in Section~4.}

The structure of this paper is as follows. Section~2 provides some background, reviewing the standard decoupling argument for D3-branes and contrasting it with cases where decoupling fails or is marginal, such as D6-branes. \textcolor{black}{Building on these benchmark examples, we identify the criteria that we will use to test decoupling.} Section~3 contains the main results of this paper: a general decoupling analysis for AdS flux branes, including expressions for the absorption probability and cross-section of perturbations originating from the branes. Section~4 is primarily illustrative, applying these general results to several examples of AdS flux vacua, including DGKT vacua, scale-separated AdS$_3$ vacua, the simplest realization of KKLT, and vacua with Romans mass. We also use these examples to discuss the effective dimensionality of DGKT and its relation to the structure of ordinary sphere compactifications. We conclude in the final section.

\newpage

\section{Strategy for testing decoupling}
When deriving a \textcolor{black}{superconformal field theory (SCFT)} from a stack of branes, it is essential to check that the theory living on the branes decouples from the bulk. In the standard examples, this leaves an SCFT on the branes decoupled from the \emph{asymptotically flat bulk}, while in the dual description the near-horizon gravitational AdS theory similarly decouples from the asymptotic region. Importantly, the decoupling analysis is performed in the low-energy limit, $E\ll M_s$, where the massive string excitations become parametrically heavy and the remaining light open-string degrees of freedom on the branes give rise to the worldvolume CFT. 
This decoupling analysis was first carried out for D3-branes by Klebanov \cite{Klebanov:1997kc}. Following a similar setup, we will probe the brane geometry with minimally coupled scalar $s$-waves and study their behaviour as their energy is taken to zero.

 The s-wave ansatz takes the form
\begin{aleq}
    \Phi(r,t)=\phi(r)e^{-i\omega t},
\end{aleq}
and we consider the low-frequency limit $\omega \rightarrow 0$.

The first way to probe decoupling is to consider an effective potential for the waves. For a D$p$-brane background, the radial equation takes the form
\begin{aleq}
    \frac{1}{r^{8-p}}\partial_r\left(r^{8-p}\partial_r\phi\right)
    +\omega^2 H_p(r)\phi=0,
    \qquad
    H_p(r)=1+\left(\frac{R}{r}\right)^{7-p}.
\end{aleq}
Introducing the dimensionless variable
\begin{aleq}
    \rho=\omega r,
    \qquad
    \phi(\rho)=\rho^{-\frac{8-p}{2}}\psi(\rho),
\end{aleq}
to rewrite the wave equation as
\begin{aleq}
    \left[-\frac{d^2}{d\rho^2}+V_{\rm eff}(\rho)\right]\psi(\rho)=0,
\end{aleq}
with
\begin{aleq}
    V_{\rm eff}(\rho)
    =
    \frac{(8-p)(6-p)}{4\rho^2}
    -1
    -\frac{(\omega R)^{7-p}}{\rho^{7-p}}.
\end{aleq}

\begin{figure}[ht]
    \centering
    \includegraphics[width=\linewidth]{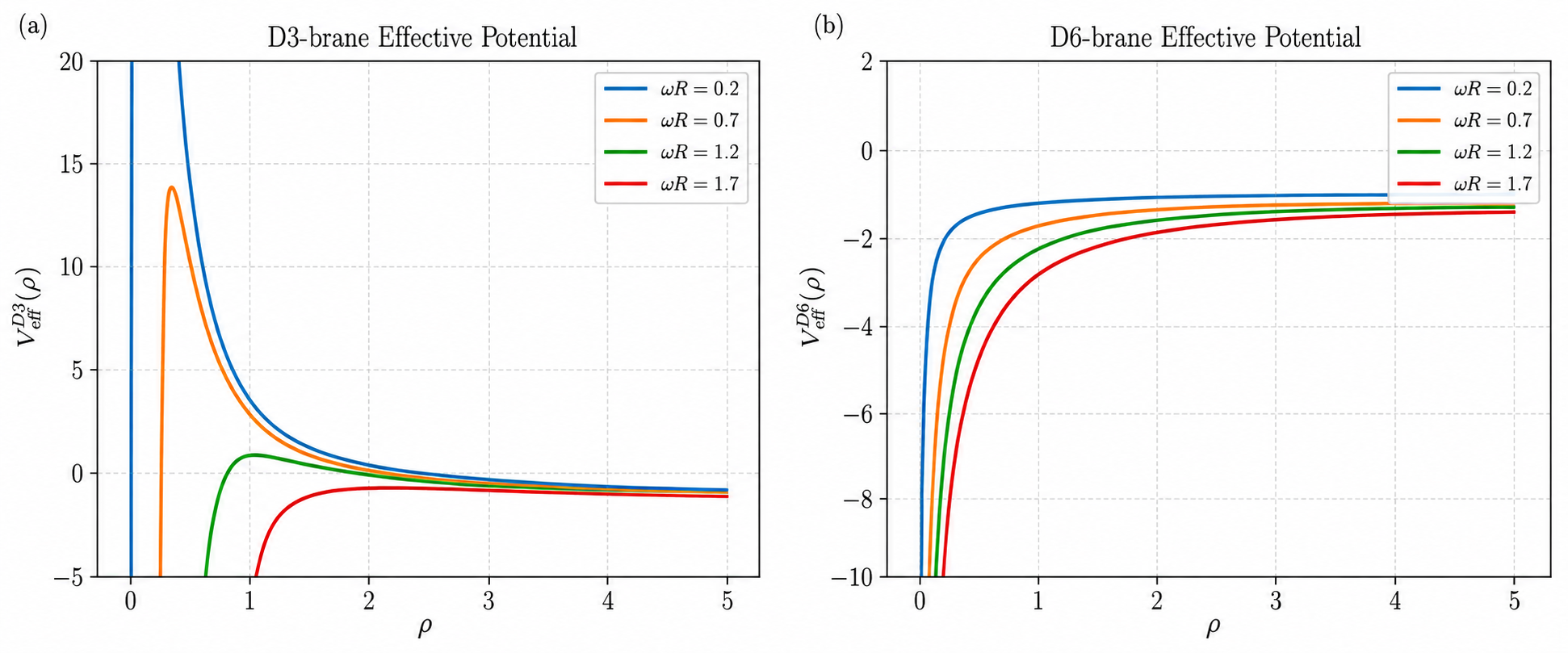}
    \caption{
    Effective potentials for scalar s-waves in D3- and D6-brane backgrounds.
    }
    \label{fig:brane-effective-potentials}
\end{figure}

For D3-branes, for instance, the effective potential takes the form
\begin{aleq}
    V_{\rm eff}^{\rm D3}(\rho)
    =
    \frac{15}{4\rho^2}
    -1
    -\frac{(\omega R)^4}{\rho^4},
\end{aleq}
and is shown in Figure~\ref{fig:brane-effective-potentials} on the left. When $\omega\rightarrow0$, there is an infinite potential barrier, signalling a successful decoupling of the D3-branes.

For D6-branes, on the other hand, whose worldvolume theory is known not to decouple, the potential takes the form
\begin{aleq}
    V_{\rm eff}^{\rm D6}(\rho)
    =
    -1
    -\frac{\omega R}{\rho},
\end{aleq}
and is shown in Figure~\ref{fig:brane-effective-potentials} on the right. There is no barrier in this case.
The second way is to perform a scattering experiment, where a wave is sent in from asymptotic infinity and the absorption probability on the branes is computed. Depending on the geometry, one can also define an absorption cross-section.

For D3-branes, the low-energy s-wave absorption probability and cross-section are
\begin{aleq}
    \mathcal P_{\rm D3}
    =
    \frac{\pi^2}{256}(\omega R)^8,
    \qquad
    \sigma_{\rm abs}^{\rm D3}
    =
    \frac{\pi^4}{8}\omega^3 R^8
    .
\end{aleq}
Both vanish in the low-energy limit $\omega R\rightarrow 0$, consistently with the decoupling of D3-branes from the bulk.

For D6-branes, absorption probability and cross-section can behave like
\begin{aleq}
    \mathcal P_{\rm D6}
    = \mathcal{O}(1), \quad \sigma_{\rm abs}^{\rm D6} \sim \omega^{-2}
\end{aleq}

Thus the absorption probability does not vanish, while the cross-section diverges.

As an interesting limiting case, we mention brane set-ups whose near-horizon geometry contains an $\mathrm{AdS}_2$ factor. For example, for an M2-M2-M5-M5 intersection \cite{Klebanov:1996mh} giving a four-dimensional black hole with near-horizon geometry
\begin{aleq}
    \mathrm{AdS}_2\times S^2,
\end{aleq}
one finds
\begin{aleq}
    \mathcal P_{\rm M2M2M5M5}
    =
    4\omega^2 R_{S^2}^2, \quad \mathcal \sigma_{\rm M2M2M5M5}
    =
    4\pi R_{S^2}^2    
\end{aleq}
Thus the absorption probability vanishes, but the absorption cross-section approaches a constant. This is the standard low-frequency theorem for black holes \cite{Das:1996we, Emparan:1997rt}: the scalar absorption cross-section approaches the horizon area. In two dimensions, the corresponding domain-wall geometries behave as black holes. We will therefore say that these AdS$_2$ vacua are \emph{weakly decoupling}, in contrast with \emph{strongly decoupled} systems, for which the absorption cross-section also vanishes.

The downside of the effective-potential method is that the presence of an infinite potential barrier in a particular coordinate system is only a \emph{sufficient} condition for weak-type decoupling. The absorption probability and absorption cross-section, on the other hand, are coordinate-independent, and their vanishing gives \emph{necessary} conditions for decoupling. However, defining these quantities requires a scattering experiment, which usually relies on an asymptotically flat spacetime. For scale-separated AdS vacua, there is no Minkowski asymptotic infinity. Nevertheless, we can still perform an analogous scattering experiment by considering an \emph{emission} probability. Its computation is equivalent to that of an absorption probability whenever the asymptotic solutions to the wave equation remain oscillatory. This is the probability for a perturbation sourced near the branes to be emitted to a large radial distance. We will use the terminology \emph{absorption} and \emph{emission} probabilities interchangeably, because in cases where both quantities can be defined, they are equal. We summarise the different decoupling criteria in Table~\ref{tab:decoupling-diagnostics}.

\begin{table}[t]
\centering
\begin{tabular}{c|c|p{6.5cm}}
\hline
Diagnostic & Condition & \centering Interpretation \tabularnewline
\hline
Infinite potential barrier
& $V_{\rm eff}\rightarrow \infty$
& Sufficient for weak decoupling \\

\hline
Absorption probability
& $\mathcal P \rightarrow 0$
& Necessary and sufficient for weak decoupling \\

\hline
Absorption cross-section
& $\sigma_{\rm abs}\rightarrow 0$
& Necessary and sufficient for strong decoupling \\
\hline
\end{tabular}

\caption{Summary of the diagnostics for decoupling.}
\label{tab:decoupling-diagnostics}
\end{table}
\newpage

\section{Decoupling of branes for AdS flux vacua}

In this section, we study perturbations on brane geometries sourcing general
AdS flux vacua \cite{Apers:2025pon,Apers:2026lgi,Bedroya:2025ltj}. We are
interested in the probability for a perturbation to reach large radial
distances, which we use as a measure of the decoupling of the brane from the
bulk.

\begin{figure}[ht!]
    \centering
    \includegraphics[width=0.9\linewidth]{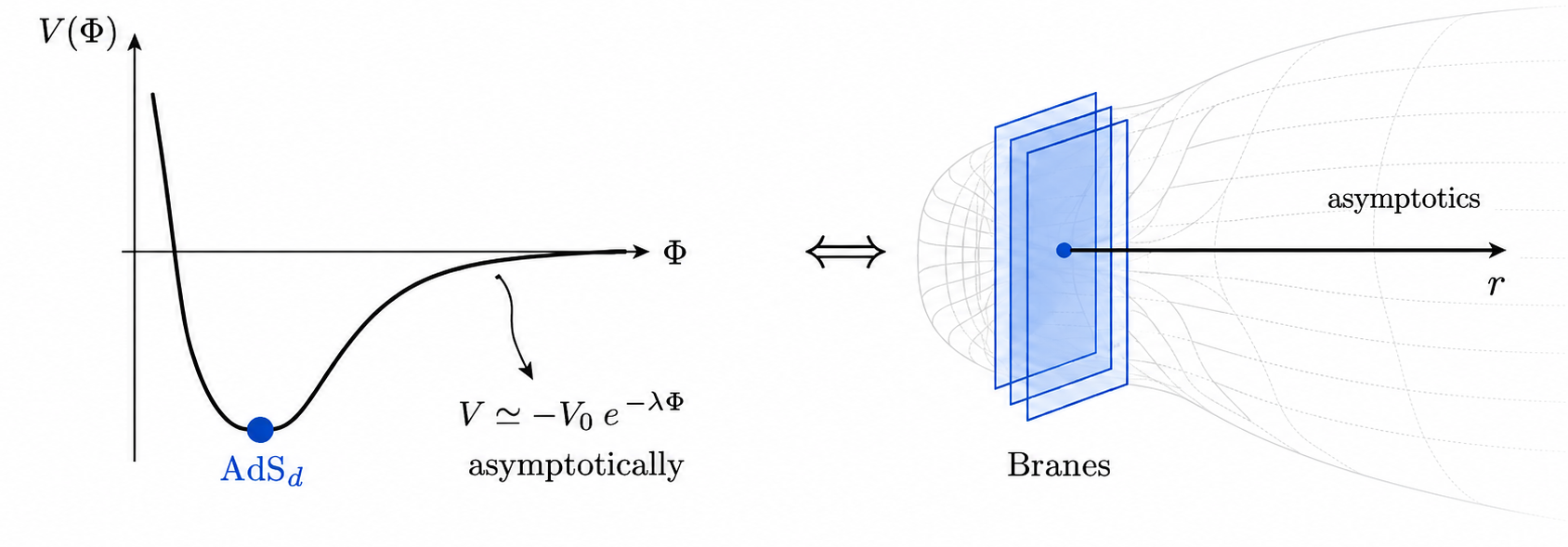}
    \caption{Equivalence between a scalar potential with an $\mathrm{AdS}_d$ vacuum and an asymptotically exponential tail, and a brane construction with radial direction $r$. }
    \label{fig:placeholder}
\end{figure}

Interestingly, for this purpose, we do not need to know the full brane geometry (as in \cite{Apers:2026lgi}). The only relevant
data are the dimension \(d\) of the AdS vacuum and the steepness \(\lambda\)
of the asymptotic scalar potential in a one-modulus truncation, see Figure \ref{fig:placeholder}.

We assume that the geometry interpolates between an \(\mathrm{AdS}_d\) throat,
\begin{equation}
    ds_d^2
    =
    \frac{L_{\rm AdS}^2}{z^2}
    \left(
        dz^2 + ds_{d-1,\rm flat}^2
    \right),\label{eq:AdSmetric}
\end{equation} 
and an asymptotic region governed by
\begin{equation}
    V(\Phi)
    =
    -V_0 e^{-\lambda\Phi},
    \qquad
    V_0>0,
    \qquad
    \lambda>0,
\end{equation}
where \(\Phi\) is canonically normalised.

In terms of the proper radial coordinate \(r\), the asymptotic metric takes
the form
\begin{equation}
    \label{eq:domain-wall-metric}
    ds_d^2
    =
    dr^2
    +
    e^{2A(r)}ds_{d-1,\rm flat}^2,
    \qquad
    A(r)=q\log r.
\end{equation}
The exponent \(q\) is determined by \(\lambda\),
\begin{equation}
    q =
    \begin{cases}
    \dfrac{4}{(d-2)\lambda^2},
    & \lambda < 2\sqrt{\dfrac{d-1}{d-2}}, \\[1.2em]
    \dfrac{1}{d-1},
    & \lambda \geq 2\sqrt{\dfrac{d-1}{d-2}} .
    \end{cases}
\end{equation}

\textcolor{black}{Note that the asymptotic scaling geometries considered here are closely related to the dynamical cobordisms in \cite{
Buratti:2021dynamical,
Buratti:2021cobordism,
Angius:2022local,
Blumenhagen:2023cobordism}. If extended in the opposite radial direction, the scaling solution would end at a finite-distance ETW singularity, whereas in our brane geometries this endpoint is replaced by the brane-sourced AdS throat.}

As a probe, we consider a minimally coupled massless scalar. We solve the wave
equation in the AdS throat and, whenever possible, match it to the asymptotic
solution. When the asymptotic region does not support propagating modes, the
effective potential is enough to determine the decoupling behaviour.

The form of the asymptotic solution depends on how steep the potential is, so we consider two cases. For $q<1$, or equivalently $\lambda > \frac{2}{\sqrt{d-2}}$, the asymptotic solutions are oscillatory and can be matched to the near-region solutions. For shallower potentials, with $q>1$ (equivalently $\lambda < \frac{2}{\sqrt{d-2}}$), the asymptotic solutions are no longer oscillatory. In this case, we cannot define an absorption probability in the same way, and instead turn to the effective potential.

\subsection{Near-region solution}
For a minimally coupled massless scalar in \(\mathrm{AdS}_d\)\textcolor{black}{, described by the metric \eqref{eq:AdSmetric},} the equation of motion for an s-wave mode $\phi(z) e^{-i\omega t}$  is given by
\begin{align}
    0=z^2 \phi''(z) - (d-2)\,z\,\phi'(z) + (\omega z)^2 \phi(z),
\end{align}
with the ingoing solution in the \(\mathrm{AdS}_d\) throat describing the near region given by
\begin{equation}
    \phi_{\rm near}(z)
    =
    C z^\alpha H^{(1)}_\alpha(\omega z),
    \qquad
    \alpha \equiv \frac{d-1}{2}.
\end{equation}
In the limit $\omega z \ll 1 $, the solution becomes dominated by the zero mode. Using the small argument expansion for the Hankel function, the solution takes the form 
\begin{equation}
    \phi_{\rm near}(z)
    =
    C\,
    \frac{2^\alpha \Gamma(\alpha)}{i\pi  \omega^{\alpha}}
    + \mathcal O(z^{2\alpha}) .
    \label{eq:near-overlap-exp-potential}
\end{equation}
The leading term is radial-independent and must be matched to the constant mode of the asymptotic solution.

\subsection{The case \texorpdfstring{$q<1$}{q<1} }
\subsubsection{Far-region solution}
For a minimally coupled massless scalar, we consider an s-wave mode $\phi(r) e^{-i\omega_\infty t_\infty}$ in the asymptotic region described by the metric in Eq.\eqref{eq:domain-wall-metric}, where $t_\infty$ denotes the asymptotic time coordinate and $\omega_\infty$ its conjugate frequency, related to the corresponding near-region quantities by $\omega_\infty t_\infty = \omega t $. The corresponding equation of motion is then
\begin{equation}
    \phi''(r)
    +
    \frac{(d-1)q}{r}\phi'(r)
    +
    \omega^2_{\infty} r^{-2q}\phi(r)
    =
    0 .
    \label{eq:far-wave-exp-potential}
\end{equation}
 For \(q<1\), define
\begin{equation}
    b \equiv 1-q,
    \qquad
    \xi \equiv \frac{\omega_{\infty}}{b} r^b,
    \qquad
     \nu \equiv \frac{q (d - 1)-1}{2b}.
\end{equation}
In terms of these parameters, the wave equation in the far-region becomes
\begin{equation}\label{eq:far-wave-eqn-xi}
    \phi''(\xi)+\frac{(d-2)q}{(1-q)\xi}\phi'(\xi)+\phi(\xi)=0,
\end{equation}
with its general solution given by
\begin{equation}
    \phi_{\rm far}(\xi)
    =
    \xi^{-\nu}
    \left[
    \mathcal A J_\nu(\xi)
    +
    \mathcal B Y_\nu(\xi)
    \right].
    \label{eq:far-bessel-solution-unified}
\end{equation}
In the limit \(\xi\ll 1\), the first branch behaves like
\begin{align}
    \xi^{-\nu}J_\nu(\xi)
    &=\frac{2^{-\nu}}{\Gamma(\nu+1)}
    +\mathcal{O}(\xi^2),
\end{align}
whereas the \(Y_\nu\) branch is singular for \(\nu>0\) and logarithmic for \(\nu=0\). 
\subsubsection{Overlap region}
There is a low-frequency regime in which both solutions reduce to the universal zero-frequency mode. This happens in the region where $\omega z \ll 1$ and \textcolor{black}{$\omega_{\infty}  r^b \ll 1$} are both true, where $\omega_{\infty}=c_{\infty}\omega\sim \omega$. To compare these inequalities one must introduce a common radial coordinate $u$. Here we will choose it such that the $(d-1)$-dimensional metric reduces in both the near and far regions to $u^2 ds^2_{d-1,\rm flat}$, which is satisfied under the identifications 
\begin{align}
    z \equiv \frac{L_{\rm AdS}}{u}, \hspace{2 em} r\equiv u^{1/q}.
\end{align}
Therefore, the regions $\omega z= \omega L_{\rm AdS}/u \ll 1$ and $\omega  r^b =\omega  u^{b/q} \ll 1$ can have overlap if 
\begin{align}
    \omega L_{\rm AdS}^b \ll1.
\end{align}
\textcolor{black}{Matching the near and far solutions, in Eq. \eqref{eq:near-overlap-exp-potential} and Eq. \eqref{eq:far-bessel-solution-unified} respectively,  to leading order in the low frequency regime defined above} sets
\begin{align}
    \mathcal A_{\rm LO}
    =
    C\,
    \frac{
    2^{\alpha+\nu}
    \Gamma(\alpha)
    \Gamma(\nu+1)
    }{i\pi \omega^{\alpha}},\hspace{2 em} 
    \mathcal B_{\rm LO}=0.
    \label{eq:A-matching-unified}
\end{align}
On the other hand, for large \(\xi\),
\begin{equation}
    J_\nu(\xi)
    \simeq
    \sqrt{\frac{2}{\pi\xi}}
    \cos\left(
    \xi-\frac{\pi\nu}{2}-\frac{\pi}{4}
    \right),
\end{equation}
and so, the far solution contains equal incoming and outgoing waves. Selecting the incoming part, the solution in the far-region after matching becomes 
\begin{align}
    \phi_{\rm far}^{\rm in}(\xi) = C\,
    \frac{
    2^{\alpha+\nu}
    \Gamma(\alpha)
    \Gamma(\nu+1)
    }{i\pi \omega^{\alpha}}\cdot \frac{1}{2}H_\nu^{(1)}(\xi).
    \label{eq:far-overlap}
\end{align}
\subsubsection{Absorption probability}
Given the solution in Eq.\eqref{eq:near-overlap-exp-potential} for the scalar wave equation in the near region, the corresponding absorbing flux into the throat is 
\begin{equation}
    \mathcal F_{\rm near}
    =
    \frac{1}{2i} \sqrt{-g}\,g^{zz}
    \left(
    \phi^*\partial_z\phi
    -
    \phi\partial_z\phi^*
    \right)
    =\frac{2L_{\rm AdS}^{d-2}}{\pi}|C|^2 .
    \label{eq:ads-absorbing-flux}
\end{equation}
Similarly, for the solution in Eq.\eqref{eq:far-overlap} in the far region, the corresponding incoming flux at infinity
\begin{equation}
    \mathcal F^{\rm in}_{\rm far}
    =
    \frac{1}{2i}
    \sqrt{-g}\,g^{rr}
    \left(
    \phi^*\partial_r\phi
    -
    \phi\partial_r\phi^*
    \right) =  \frac{b|\mathcal A|^2}{2\pi \,c_{\infty}^{d-1}}
    \left(
    \frac{b}{c_{\infty}\omega}
    \right)^{2\nu},
\end{equation}
where we have introduced the coefficient \textcolor{black}{$c_\infty \equiv t/t_\infty$} accounting for a different normalisation of the time coordinates in the metrics presented for the near and far regions, such that the factor $\sqrt{-g}$ is normalised in the same way as in the near region. It enters the normalisation of the frequencies as well,  $\omega_\infty =c_\infty \omega$.

Substituting the matched coefficient \eqref{eq:A-matching-unified}, we obtain
\begin{equation}\label{eq:flux-far-qless1}
    \mathcal F^{\rm in}_{\rm far}
    =
    \frac{b}{2\pi c_\infty^{2\alpha + 2\nu} }
    \left(
    \frac{b}{\omega}
    \right)^{2\nu}
    |C|^2
    \left[
    \frac{
    2^{\alpha+\nu}
    \Gamma(\alpha)
    \Gamma(\nu+1)
    }{\pi}
    \right]^2
    \omega^{-2\alpha}.
\end{equation}
As a result, the leading absorption probability $ \mathcal P_{\rm abs}\equiv (\mathcal F_{\rm near} /\mathcal F^{\rm in}_{\rm far})$ is
\begin{equation}
    \mathcal P_{\rm abs}^{\rm LO}
    =
    \frac{
    2^{2-2(\alpha+\nu)}\pi^2
    c_\infty^{2\alpha + 2\nu}}{
    b^{2\nu+1}
    \nu^2 \Gamma(\alpha)^2
    \Gamma(\nu)^2
    }
    \,
   (\omega  L_{\rm AdS}^b)^{2(\alpha + \nu) }, 
    \label{eq:Pabs-unified-alpha-nu}
\end{equation}
where recall $\alpha \equiv (d-1)/2$ and $\nu \equiv (q(d-1)-1)/(2(1-q))$. Consequently, the absorption probability scales with the frequency as 
\begin{equation}
    \mathcal P_{\rm abs}^{\rm LO}
    \sim \omega^\Delta, \quad \Delta = 2(\alpha + \nu).
    \label{eq:Pabs-scaling}
\end{equation}
For the exponential potential regime,
\[
    \frac{2}{\sqrt{d-2}}
    <
    \lambda
    <
    2\sqrt{\frac{d-1}{d-2}},
\]
where the lower bound corresponds to restricting to the case considered in this section, $q < 1$, one has
\begin{equation}
    q=\frac{4}{(d-2)\lambda^2},
    \qquad
    \Delta =
    \frac{(d-2)^2\lambda^2}{(d-2)\lambda^2-4}.
\end{equation}
In this regime, $\Delta >d-1$ and the absorption probability vanishes as $\omega \rightarrow 0$. 

The very steep, or kination, regime where
\begin{aleq}
    \lambda \geq  2\sqrt{\frac{d-1}{d-2}}
\end{aleq}
corresponds to
\begin{equation}
    q=\frac{1}{d-1},
    \qquad
    \Delta = d-1.
\end{equation}
Therefore, the absorption probability vanishes again in the low-energy
limit, $\omega \rightarrow 0$.


\subsubsection{Absorption cross-section}

We can go one step further and determine the scaling of the absorption \emph{cross-section}. To do so, we first need to identify the effective dimension in which the asymptotic waves propagate. For branes in asymptotically flat spacetime, this is simply the number of transverse spatial dimensions; for example, a D3-brane has six transverse dimensions. For the non-flat asymptotics considered here, however, there is no analogous literal geometric interpretation. Nevertheless, the asymptotic wave equation, as given in Eq.~\eqref{eq:far-wave-eqn-xi}, is identical to that of a wave propagating in flat spacetime with an effective, and in general non-integer, number of transverse dimensions. We therefore use this effective dimensionality to define the corresponding absorption cross-section.

In the regime
\begin{equation}
\frac{2}{\sqrt{d-2}}
<
\lambda
<
2\sqrt{\frac{d-1}{d-2}},
\end{equation}
the effective transverse dimension is
\begin{equation}
d_{\rm eff}
=
1+
\frac{4(d-2)}
{(d-2)\lambda^2-4}.
\end{equation}
In the very steep, kination-like regime, on the other hand, one finds
\begin{equation}
d_{\rm eff}=2.
\end{equation}
The absorption cross-section is then obtained by dividing the absorption probability by the appropriate asymptotic phase-space factor,
\begin{equation}
\sigma_{\rm abs}
\sim
\frac{\mathcal{P}_{\rm abs}}
{\omega^{d_{\rm eff}-1}}.
\end{equation}
Using the low-frequency scaling of $\mathcal{P}_{\rm abs}$ found in Eq.~\eqref{eq:Pabs-scaling} in terms of the effective transverse dimension, $\mathcal{P}_{\rm abs} \sim \omega ^{d+d_{\rm eff}-3}$, both regimes then lead to the universal behaviour
\begin{equation}
\sigma_{\rm abs}
\sim
\omega^{d-2}.
\end{equation}
The resulting power is independent of the steepness parameter $\lambda$ and depends only on the dimension $d$ of the AdS spacetime.

\subsection{The case \texorpdfstring{$q>1$}{q>1} }
For $q>1$, or equivalently $\lambda<2/\sqrt{d-2}$, the asymptotic wave equation is again given by Eq.~\eqref{eq:far-wave-exp-potential}. As in
Section~3.2.1, it can be written in Bessel form by defining
\begin{equation}
    b \equiv q-1,
    \qquad
    \xi \equiv \frac{\omega_\infty}{b}r^{-b},
    \qquad
    \nu \equiv \frac{q(d-1)-1}{2b},
\end{equation}
so that
\begin{equation}
    \phi_{\rm far}(\xi)
    =
    \xi^\nu\left[A J_\nu(\xi)+B Y_\nu(\xi)\right].
\end{equation}
The crucial difference with the $q<1$ case is that now
$\xi\to 0$ as $r\to\infty$, rather than $\xi\to\infty$ as before. Using the
small-argument expansion of the Bessel functions therefore gives
\begin{equation}
    \phi_{\rm far}(r)
    =
    \phi_0+\phi_1 r^{1-(d-1)q}+\ldots .
\end{equation}
where $\phi_0$ and $\phi_1$ are constants. Hence the asymptotic solutions are non-oscillatory and do not admit the usual decomposition into incoming and outgoing waves. The standard scattering interpretation is therefore not directly applicable, and we instead turn to the effective potential.

Defining
\begin{equation}
    \phi(r)
    =
    r^{-\frac{(d-1)q}{2}}\,u(r),
\end{equation}
the radial equation becomes
\begin{equation}
    -u''(r)+V_{\rm eff}(r)u(r)=0,
\end{equation}
with
\begin{equation}
    V_{\rm eff}(r)
    =
    -\omega^2 r^{-2q}
    +
    \frac{A}{r^2},
    \qquad
    A
    =
    \frac{(d-1)q\bigl((d-1)q-2\bigr)}{4}>0.
\end{equation}
Since \(q>1\), the first term decays faster than \(1/r^2\), so
\begin{equation}
    V_{\rm eff}(r)
    \sim
    \frac{A}{r^2},
    \qquad r\to\infty.
\end{equation}
The asymptotic potential is therefore positive, consistent
with the absence of propagating modes at infinity. Hence no outgoing flux
reaches the asymptotic region, and
\begin{aleq}
    \mathcal P_{\rm em}=0.
\end{aleq}

\subsection{The case \texorpdfstring{$q=1$}{q=1} }
\textcolor{black}{
For $q=1$, or equivalently $\lambda=2/\sqrt{d-2}$,  the asymptotic scalar equation given in Eq. \eqref{eq:far-wave-exp-potential} becomes
\begin{equation}
    r^2 \phi'' + (d-1) r \phi' + \omega^2 \phi = 0.
\end{equation}
Introducing
\begin{equation}
    \phi(r) = r^{-\frac{d-2}{2}}u(r),
    \qquad x = \log r,
\end{equation}
the equation becomes
\begin{equation}
    -u''(x)
    + \frac{(d-2)^2}{4}u(x)
    = \omega^2 u(x).
\end{equation}
The asymptotic region therefore supports oscillatory incoming and outgoing modes only for
\begin{equation}
    \omega > \omega_{\rm crit},
    \qquad
    \omega_{\rm crit} = \frac{d-2}{2},
\end{equation}
in the normalisation of Eq.~(3.3). Below this threshold, imposing the decaying canonical branch at infinity gives zero asymptotic flux and hence a vanishing emission probability,
\begin{equation}
    \mathcal{P}_{\rm em} = 0,
    \qquad
    \omega < \omega_{\rm crit}.
\end{equation}
}
\subsection{Summary}
\begin{table}[t]
    \centering
    \renewcommand{\arraystretch}{1.35}
    \begin{tabular}{c c c c}
        \toprule
        Steepness & $\mathcal{P}_{\rm em}$ & $d_{\rm eff}$ &
        $\sigma_{\rm abs}$ \\
        \midrule
        $\displaystyle
        \lambda \geq 2\sqrt{\frac{d-1}{d-2}}
        $
        &
        $\omega^{d-1}$
        &
        $2$
        &
        $\omega^{d-2}$
        \\[2mm]

        $\displaystyle
        \frac{2}{\sqrt{d-2}}
        <\lambda<
        2\sqrt{\frac{d-1}{d-2}}
        $
        &
        $\omega^{d+d_{\rm eff}-3}$
        &
        $\displaystyle
        1+\frac{4(d-2)}{(d-2)\lambda^2-4}
        $
        &
        $\omega^{d-2}$
        \\[3mm]

        $\displaystyle
        \lambda=\frac{2}{\sqrt{d-2}}
        $
        &
        $0\quad (\omega<\frac{d-2}{2})$
        &
        -- & -- \\[1mm]

        $\displaystyle
        \lambda<\frac{2}{\sqrt{d-2}}
        $
        &
        $0$
        &
        -- & -- \\
        \bottomrule
    \end{tabular}
    \caption{Low-energy emission probability for the different asymptotic
regimes.
    \label{tab:emission-regimes}}
\end{table}

In this section, we determined the probability for a perturbation on the
brane to escape to radial infinity for branes sourcing general AdS flux
vacua. The only input data are the dimension \(d\) of the AdS
vacuum and the steepness \(\lambda\) of the asymptotic potential in a
one-modulus truncation.

We find that, for all such AdS vacua, the emission probability is suppressed
at least in the low-energy limit, \(\omega\to 0\). In this sense, the brane
decouples from the asymptotic bulk. Moreover, the decoupling becomes stronger
as the asymptotic potential becomes shallower. For sufficiently shallow
potentials, emission to infinity vanishes for all frequencies. The different
regimes are summarized in Table~\ref{tab:emission-regimes}.

As noted in \cite{Bedroya:2025ltj}, scale-separated AdS vacua tend to be associated
with steep asymptotic potentials. According to our analysis, such vacua
therefore exhibit a weaker form of decoupling, although the low-energy
emission probability still vanishes, as required for a holographic
description.

\textcolor{black}{
Note that the regime $\lambda \leq 2/\sqrt{d-2}$, where decoupling appears
stronger, corresponds to accelerated expansion in the associated
positive-potential cosmology \cite{Rudelius:2022}, under the domain-wall/cosmology correspondence\footnote{For example, see \emph{double Wick rotation} in \cite{Bedroya:2025ltj}},
which reverses the sign of the potential. The other threshold,
$\lambda = 2\sqrt{(d-1)/(d-2)}$, also resonates with bounds on critical
exponents arising in dynamical cobordism criteria~\cite{CalderonInfante:2026,Makridou:2026}.}

\newpage
\section{Examples}
\subsection{Classical AdS\texorpdfstring{$_4$}{4} vacua with scale separation}

Known constructions of scale separation in AdS$_4$ using classical ingredients are the DGKT-CFI vacua \cite{DeWolfe:2005uu,Camara:2005dc} in massive IIA, and the related constructions in IIA without Romans mass in \cite{Cribiori:2021djm}.

The brane dual for the DGKT vacua and their decoupling was discussed in \cite{Kounnas:2007dd,Apers:2025pon,Apers:2026lgi}. The ten-dimensional metric of the brane dual is given by \cite{Apers:2025pon,Apers:2026lgi}
\begin{aleq}\label{DGKT_full}
    ds_{10}^2
    &=
    H(r)^{3/2}\, dr^2
    + r^{-10/9} H(r)^{-3/2}\, ds^2_{3,\text{non-compact}}
    + r^{2/3} H(r)^{1/2}\, ds^2_{6,\text{compact}},
    \\
    e^{\phi}
    &=
    r^{-1} H(r)^{-3/4},
    \qquad
    H(r)
    =
    1 + \frac{R^{4/3}}{r^{4/3}},
\end{aleq}
where $R$ is a constant. A massless scalar s-wave in this background obeys
\begin{equation}
    \phi''(r)
    + \frac{7}{3}\,\frac{\phi'(r)}{r}
    + \omega^{2}\,
    \frac{\bigl(R^{4/3} + r^{4/3}\bigr)^{3}}{r^{26/9}}
    \phi(r)
    = 0.
\end{equation}
Upon redefining the scalar $\psi(r)=r^{7/6}\phi(r)$, the wave equation can be written directly in the following form as
\begin{aleq}
    \left[-\frac{d^2}{dr^2}+V_{\rm DGKT}(r)\right]\psi(r)=0,
    \qquad
    V_{\rm DGKT}(r)
    =
    -\omega^2
    \frac{\bigl(r^{4/3}+R^{4/3}\bigr)^3}{r^{26/9}}
    +\frac{7}{36r^2}.
    \label{DGKTpotbar}
\end{aleq}

\begin{figure}[t]
    \centering
    \includegraphics[width=0.7\linewidth]{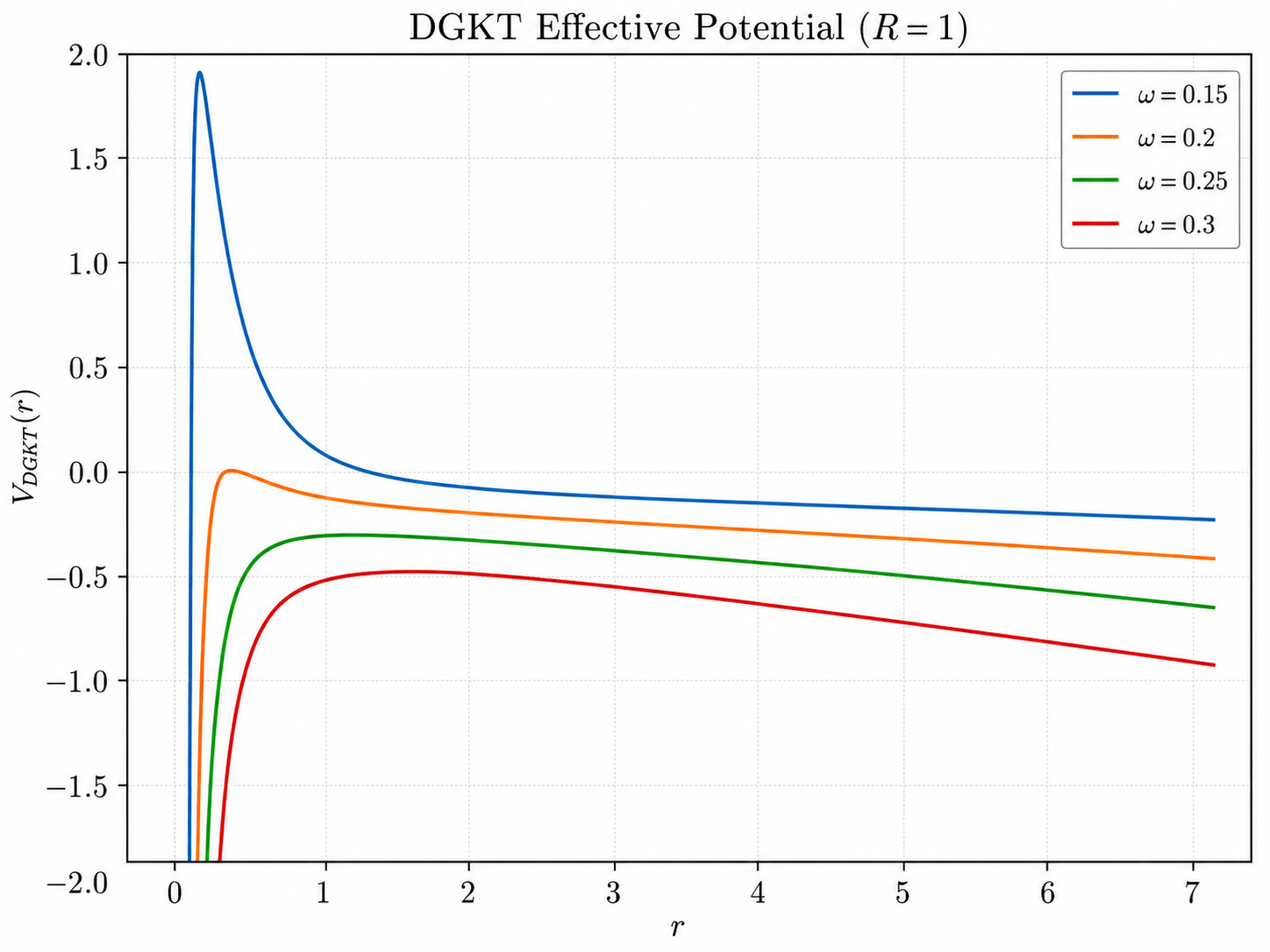}
    \caption{
    Effective potentials for scalar s-waves in DGKT brane backgrounds.
    }
    \label{fig:DGKTpotbar}
\end{figure}

The effective potential is shown in Figure~\ref{fig:DGKTpotbar}. In the low-frequency limit it develops an infinite potential barrier separating the near-horizon region from the asymptotic bulk. By the criterion discussed in Section~2, this is sufficient to conclude that the DGKT branes decouple from the bulk in the low-energy limit. In contrast to the Freund--Rubin vacua, the potential does not approach a constant at infinity.

We can obtain the absorption probability directly from the general analysis of Section~3.2.3, without needing to know the full brane geometry. For this purpose, the only additional information required is the steepness of the scalar potential in the asymptotic region of moduli space.
In terms of the two moduli $s$ and $u$, the DGKT scalar potential takes the schematic form
\begin{equation}
    V(s,u)
    =
    \frac{1}{s^3}
    \left[
        \frac{A_{F_4}}{us}
        +
        A_{F_0}\frac{u^3}{s}
        +
        A_{H_3}\frac{s}{u^3}
        -
        A_{O6}
    \right].
        \label{eq:DGKT-potential}
\end{equation}
In the asymptotic large-moduli region followed by the brane solution, the $F_4$ contribution is subleading, while the leading terms arise from the Romans mass, the $H_3$ flux, and the O6-plane contribution. Along this asymptotic trajectory, we have the one-modulus truncation
\begin{equation}
    s=u^3,
        \label{eq:DGKT-truncation}
\end{equation}
for which the potential scales as
\begin{equation}
    V_{\rm asym}\sim s^{-3}\sim u^{-9}.
\end{equation}
The kinetic terms along this trajectory define the canonically normalised field
\begin{equation}
    \Phi=\sqrt{\frac{39}{2}}\,\log u,
        \label{eq:DGKT-canonical-modulus}
\end{equation}
and hence the asymptotic potential can be written in the exponential form
\begin{equation}
    V_{\rm asym}\sim -V_0 e^{-\lambda\Phi},
    \qquad
    \lambda^2=\frac{54}{13}.
\end{equation}

We can now immediately apply the result of Section~3.2.3,
\begin{equation}
    \mathcal P_{\rm abs}
    \sim \omega^\Delta,
    \qquad
    \Delta
    =
    \frac{(d-2)^2\lambda^2}
    {(d-2)\lambda^2-4}.
\end{equation}
For the DGKT vacua,
\begin{equation}
    d=4,
    \qquad
    \lambda^2=\frac{54}{13}
    \qquad\Longrightarrow\qquad
    \Delta=\frac{27}{7},
\end{equation}
and therefore
\begin{equation}\label{abs_dgkt}
    \mathcal P_{\rm abs}
    =
    \frac{2^{1/7}\pi}{\Gamma(10/7)^2}
    \left(\frac{27}{14}\right)^{13/7}
    c_\infty^{27/7}L_{\rm AdS}^{2}
    \,\omega^{27/7} \sim \omega^{27/7}.
\end{equation}
Thus the absorption probability vanishes in the low-energy limit, providing the coordinate-independent confirmation of the decoupling inferred from the infinite potential barrier.

For the vacua without Romans mass \cite{Cribiori:2021djm}, the corresponding brane background takes the following string frame form,
\begin{equation}
\begin{aligned}
ds^2_{10}
={}&
H(r)^{3/2}\,dr^2
+
H(r)^{-3/2}\,r^{-10/9}\,ds^2_{3,\mathrm{nc}}
\\
&+
H(r)^{-1/2}\,r^{-2/3}\,dy^2_{1,2}
+
H(r)^{1/2}\,r^{-2/3}
\left(
dy^2_{3,4}+dy^2_{5,6}
\right),
\end{aligned}
\end{equation}
where the coordinates \(y_i\) parametrize a toroidal internal space. The dilaton and harmonic function are
\begin{equation}
    e^\phi
    =
    H(r)^{-5/4}r^{-5/3},
    \qquad
    H(r)
    =
    1+\frac{R^{4/3}}{r^{4/3}} .
\end{equation}
This geometry can be obtained by applying the flux backtracking procedure with bounded fluxes and internal curvature, and by introducing D2 and D6 branes sourced by the harmonic function \(H(r)\) \cite{Apers:2026lgi}. The scalar wave equation in this background leads to the same effective potential as in the DGKT case, and gives the low-energy absorption probability
\begin{equation}
    \mathcal P_{\rm abs}\sim \omega^{27/7}.
\end{equation}
The conclusion about decoupling is therefore unchanged.
\subsubsection{Absorption Cross-Section}

In the asymptotic region, Eq.~(4.2) can be written in terms of
\begin{equation}
    \rho=\frac{9}{14}\omega r^{14/9}
\end{equation}
as
\begin{equation}
    \frac{d^2\phi}{d\rho^2}
    +\frac{13}{7\rho}\frac{d\phi}{d\rho}
    +\phi
    =
    \frac{d^2\phi}{d\rho^2}
    +\frac{d_{\rm eff}-1}{\rho}\frac{d\phi}{d\rho}
    +\phi=0,
\end{equation}
from which we identify
\begin{equation}
    d_{\rm eff}=\frac{20}{7}.
\end{equation}
Using $\mathcal{P}_{\rm abs}\sim\omega^{27/7}$, the corresponding absorption
cross-section therefore scales as
\begin{equation}
    \sigma_{\rm abs}
    \sim \frac{\omega^{27/7}}{\omega^{ d_{\rm eff-1}}}
    =\omega^2,
\end{equation}
and hence vanishes in the low-energy limit, providing a strong form of decoupling.

For these vacua, the effective dimension seen by the waves appears to
have a more direct interpretation than for generic flux vacua. Indeed,
compactifying a hypothetical $(4+13/7)$-dimensional theory of gravity
on an $S^{13/7}$ threaded by $N$ units of flux gives a scalar potential
of the form (see Appendix \ref{app:generalised-freund-rubin})
\begin{equation}
    V(\Phi)
    =
    A_{F_4}e^{-\sqrt{26/3}\,\Phi}
    -V_0e^{-\sqrt{54/13}\,\Phi},
    \qquad
    A_{F_4}\propto N^2 .
    \label{eq:fractional-sphere-potential}
\end{equation}
Remarkably, this coincides with the DGKT potential
\eqref{eq:DGKT-potential} along the truncation
\eqref{eq:DGKT-truncation}, with the canonically normalised modulus
defined in Eq.~\eqref{eq:DGKT-canonical-modulus}, upon identifying
\begin{equation}
    V_0\equiv A_{O6}-A_{F_0}-A_{H_3}.
    \label{eq:DGKT-V0}
\end{equation}
It would be interesting to understand whether this effective
fractional dimensionality has a corresponding interpretation in the
dual CFT.

\subsection{Classical AdS\texorpdfstring{$_3$}{3} vacua with scale separation}

Several classes of parametrically scale-separated AdS$_3$ vacua constructed
from classical ingredients are known \cite{Farakos:2020phe, Farakos:2025vkn, VanHemelryck:2025qok, Arboleya:2024vnp, Arboleya:2025ocb, Tringas:2025bwe, Miao:2025rgf}. We will focus on the following

\begin{itemize}
    \item Type IIB compactifications on seven-dimensional manifolds with
    co-closed $G_2$ structure \cite{VanHemelryck:2025qok}. A particularly simple example is the nilmanifold $\mathfrak n_2\simeq\mathrm{Nil}_3\times\mathbb T^4$, which preserves minimal supersymmetry and has integer conformal dimensions for the light
    scalar operators. Scale-separated solutions also exist on several other nilmanifolds, although their flux scaling and scalar spectra are less simple than for $\mathfrak n_2$ \cite{VanHemelryck:2025qok}.

    \item Massive type IIA compactifications on singular $G_2$ orientifolds
    with $F_0$, $H_3$ and an unbounded $F_4$ flux
    \cite{Farakos:2020phe}. These constructions admit minimally
    supersymmetric $\mathcal N=1$ solutions, but their scalar spectrum does
    not exhibit integer conformal dimensions.

    \item Massive type IIA compactifications on Joyce-type $G_2$ orbifolds,
    admitting a desingularization to smooth $G_2$ manifolds
    \cite{Farakos:2025vkn}. These are minimally supersymmetric
    $\mathcal N=1$ vacua involving $F_0$, $H_3$ and $F_4$ flux, and their
    scalar spectrum has integer conformal dimensions. More general $F_4$ flux choices are required for full
    moduli stabilization.
    \end{itemize}

There are more (non)-supersymmetric type IIB orientifold compactifications 
    \cite{Arboleya:2024vnp, Arboleya:2025ocb}. Some of these solutions have
    integer conformal dimensions and admit a parametric hierarchy between
    the AdS$_3$ radius and the characteristic overall size of the internal
    space. Their brane duals are discussed in detail in \cite{ArboleyaGuarinoRoldanSudano2026} and are not discussed here.

\subsubsection{Decoupling for Type IIB on the nilmanifold \texorpdfstring{$\mathfrak n_2$}{n2}}
Applying flux backtracking \cite{Apers:2025pon} and subsequently
placing the D1- and D5-branes dual to the unbounded fluxes on top of the
flux-backtracking geometry, following the strategy of
\cite{Apers:2026lgi}, we obtain a brane geometry whose near-horizon limit
reproduces the scale-separated AdS$_3$ vacuum on the nilmanifold
$\mathfrak n_2$ of \cite{VanHemelryck:2025qok}. The resulting
ten-dimensional string-frame metric and dilaton are
\begin{align}
    ds_{10}^2
    &=(HK^5)^{1/2} d\rho^2
    +(HK^5)^{-1/2}\rho^{-4/3}dx_\mu dx^\mu
    +(HK)^{1/2}\rho^{2/3}(dy_1^2+dy_2^2)
    \nonumber\\
    &\quad
    +(HK^{-1})^{1/2}
    (dy_3^2+dy_4^2+dy_5^2+dy_6^2)
    +(HK^{-3})^{1/2}\rho^{-2/3}dy_7^2,
    \label{eq:ads3-brane-metric}\\
    e^\phi&=(HK^{-5})^{1/2}\rho^{-4/3},
    \label{eq:ads3-brane-dilaton}
\end{align}
where
\begin{equation}
    H(\rho)=1+\left(\frac{R_{D1}}{\rho}\right)^{2/3},
    \qquad
    K(\rho)=1+\left(\frac{R_{D5}}{\rho}\right)^{2/3}.
    \label{eq:ads3-harmonic-functions}
\end{equation}
Details of the construction are given in Appendix~\ref{app:ads3-absorption}.

\begin{figure}[t]
    \centering
    \includegraphics[width=0.7\linewidth]{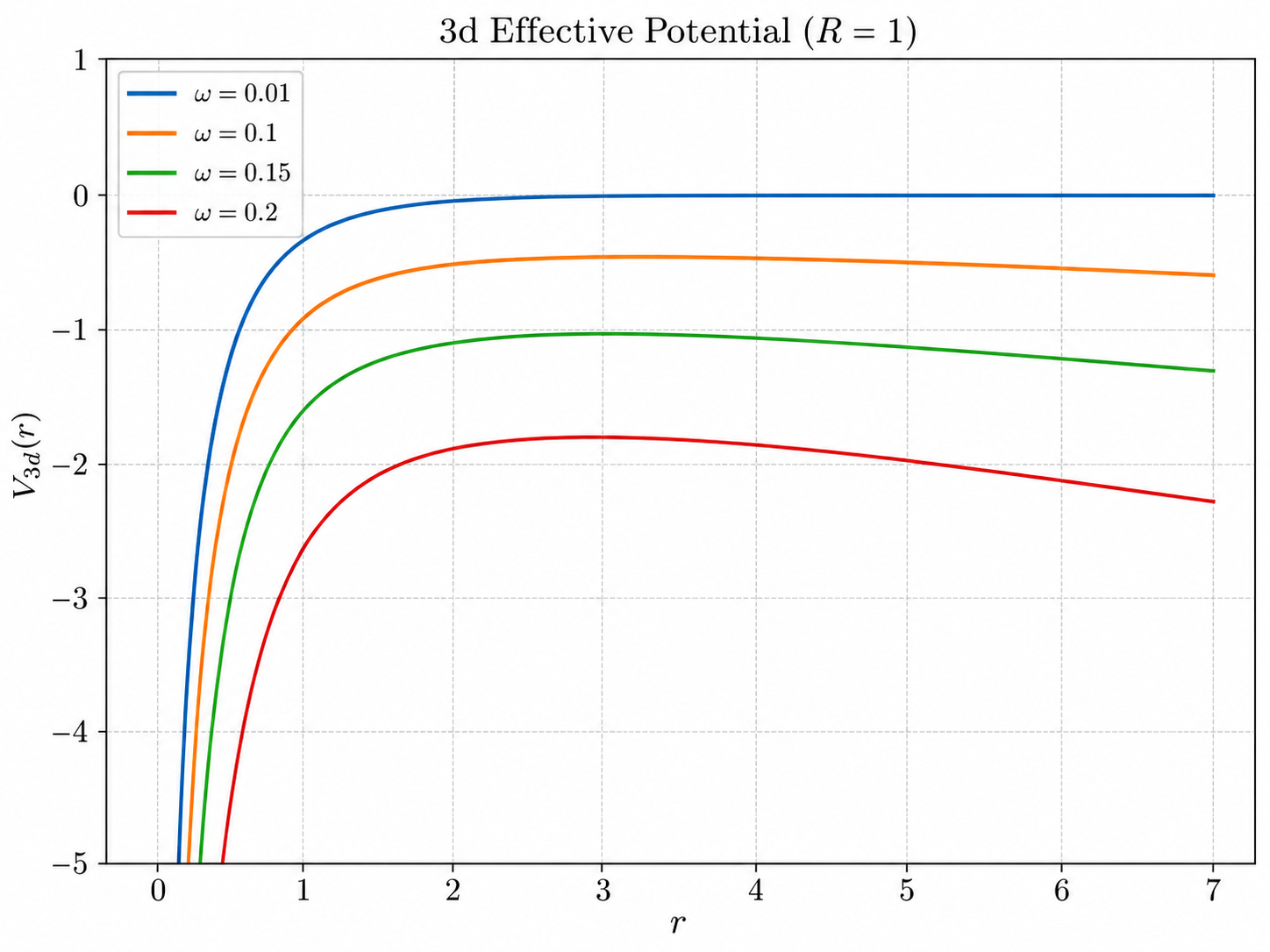}
    \caption{Effective potential for scalar $s$-waves in the
    scale-separated AdS$_3$ brane background associated with the
    D1--D5 system of Table~\ref{tab:D1D5system}.}
    \label{fig:3dpotbar}
\end{figure}

Having the full interpolating geometry allows us to derive the effective
potential for scalar waves,
\begin{equation}
    V_{\rm eff}(\rho)
    =
    -\frac{5}{36\rho^2}
    -\omega^2\rho^{4/3}
    \left[
        1+\left(\frac{R_{D1}}{\rho}\right)^{2/3}
    \right]
    \left[
        1+\left(\frac{R_{D5}}{\rho}\right)^{2/3}
    \right]^5 .
    \label{eq:ads3-effective-potential}
\end{equation}
The derivation is again given in Appendix~\ref{app:ads3-absorption}.
The potential is shown in Figure \ref{fig:3dpotbar}
and does not develop a potential barrier. Hence, in contrast
with the DGKT case, the effective potential alone is insufficient to
establish decoupling, and we must instead compute the absorption
probability.

As shown in Appendix~\ref{app:ads3-absorption}, the absorption probability behaves at low frequencies as
\begin{equation}
    \mathcal P_{\rm abs}\sim \omega^{12/5},
\end{equation}
and therefore vanishes in the low-energy limit, establishing the decoupling of the branes from the asymptotic bulk. The same result follows directly from Eq.~\eqref{eq:Pabs-scaling}, upon setting
\begin{equation}
    d=3,
    \qquad
    \lambda=4\sqrt{\frac{3}{7}},
\end{equation}
where $\lambda$ is the steepness of the asymptotic scalar potential.

We derive this asymptotic steepness in the following way.
The scalar sector of the $\mathfrak n_2$ compactification is parametrized by
the three-dimensional dilaton $s$ and the seven cycle sizes $L_i$,
$i=1,\ldots,7$. The three-dimensional dilaton is related to the
ten-dimensional string coupling by
\begin{equation}
    s^{-1}
    =
    g_s^{-1}\sqrt{L_1L_2L_3L_4L_5L_6L_7}.
    \label{eq:n2-3d-dilaton}
\end{equation}
In terms of the fields
\begin{equation}
    \phi^A=(s,L_1,\ldots,L_7),
\end{equation}
the scalar-field metric is diagonal,
\begin{equation}
    G_{AB}
    =
    \operatorname{diag}
    \left(
        \frac{2}{s^2},
        \frac{1}{2L_1^2},
        \ldots,
        \frac{1}{2L_7^2}
    \right),
    \label{eq:n2-field-space-metric}
\end{equation}
such that
\begin{equation}
    \mathcal L_{\rm kin}
    =
    -\frac{2}{s^2}(\partial s)^2
    -\frac12\sum_{i=1}^{7}
    \frac{(\partial L_i)^2}{L_i^2}.
    \label{eq:n2-kinetic-terms}
\end{equation}

Using the real superpotential $P$, 
\begin{equation}
    P
    =
    -\frac{s^3}{4\sqrt{\mathcal V_7}}\,
    \mathcal F
    +\frac{s^2\,\omega\,L_7}{4L_1L_2},
    \label{eq:n2-superpotential}
\end{equation}
with
\begin{align}
    \mathcal F
    ={}&
    F_{71}
    -F_{33}L_1L_2L_3L_4
    -F_{32}L_1L_2L_5L_6
    +F_{31}L_3L_4L_5L_6
    \nonumber\\
    &+F_{37}L_1L_3L_5L_7
    +F_{36}L_2L_4L_5L_7
    +F_{35}L_2L_3L_6L_7
    +F_{34}L_1L_4L_6L_7 .
    \label{eq:n2-flux-combination}
\end{align}

the scalar potential is
$V=G^{AB}\partial_A P\,\partial_B P-4P^2$.

Along the flux backtracking trajectory, we have the following identifications of moduli
\begin{equation}
    L_1=L_2=s^{-2/9},
    \qquad
    L_3=L_4=L_5=L_6=\text{const.},
    \qquad
    L_7=s^{2/9},
    \label{eq:n2-moduli-trajectory}
\end{equation}
so that a single canonically normalised modulus can be chosen as:
\begin{equation}
    \Phi
    =-\frac{4\sqrt{21}}{9}\log S .
    \label{eq:n2-canonical-trajectory}
\end{equation}
The potential then takes the two-exponential form
\begin{equation}
    V(\Phi)
    =
    A\,e^{-\frac{2\sqrt{21}}{3}\,\Phi}
    +
    B\,e^{-4\sqrt{3/7}\Phi},
    \label{eq:n2-potential-canonical}
\end{equation}
where
\begin{align}
A&=
    \frac14\left[
        F_{71}^{\,2}+F_{31}^{\,2}+F_{34}^{\,2}
        +F_{35}^{\,2}+F_{36}^{\,2}+F_{37}^{\,2}
    \right].\\
    B&=
    \frac14\left[
        F_{32}^{\,2}+F_{33}^{\,2}
        -2\omega(F_{32}+F_{33})+\omega^2
    \right],
\end{align}
the first term consisting of the unbounded fluxes, while the bounded fluxes arise in the second term. The asymptotic steepness sourced by the bounded ingredients is then given by
\begin{aleq}
    \lambda = 4 \sqrt{3/7}.
\end{aleq}
Moreover, the one-modulus scalar potential agrees, up to an identification
of coefficients, with that of a generalised Freund--Rubin compactification
of a hypothetical $(3+7/5)$-dimensional theory on an $S^{7/5}$, corresponding
formally to an effective $\mathrm{AdS}_3\times S^{7/5}$ geometry, see Appendix \ref{app:generalised-freund-rubin}. This suggests
that these scale-separated $\mathrm{AdS}_3$ vacua may belong to the same
generalised Freund--Rubin class as the DGKT vacua.

In this interpretation, the effective transverse dimension seen by the waves is
\begin{equation}
    d_{\rm eff}
    =
    1+\frac{7}{5}
    =
    \frac{12}{5}.
\end{equation}
Together with $P_{\rm abs}\sim\omega^{12/5}$, this gives the low-frequency
absorption cross-section
\begin{equation}
    \sigma_{\rm abs}
    \sim
    \frac{P_{\rm abs}}{\omega^{d_{\rm eff}-1}}
    \sim
    \omega .
\end{equation}
The cross-section therefore vanishes linearly in the low-energy limit,
providing a strong form of decoupling for these vacua.

\subsubsection{Decoupling for massive Type IIA on a singular \texorpdfstring{$G_2$}{G2} manifold}

For these vacua, we do not have access to the full interpolating brane
geometry. This is, however, not necessary for determining the low-energy
decoupling. Using the flux-backtracking results of
\cite{Apers:2025pon}, the asymptotic moduli trajectory is
\begin{equation}
    s(r)\sim r^{13/16},
    \qquad
    u(r)\sim r^{1/8},
\end{equation}
such that
\begin{equation}
    s\sim u^{13/2}.
\end{equation}
After removing the unbounded $F_4$ flux, the residual scalar potential is
\begin{equation}
    V
    =
    A_{F_0}u^{7/2}s^{-3}
    +
    A_{H_3}u^{-3}s^{-2}
    -
    A_{O6}u^{1/4}s^{-5/2}.
\end{equation}
Along the flux-backtracking trajectory, all three contributions scale in
the same way,
\begin{equation}
    V_{\rm asym}
    \sim
    -V_0\,u^{-16}.
\end{equation}
The corresponding canonically normalised modulus is determined as
\begin{equation}
    \Phi
    =
    2\sqrt{11}\log u
    =
    \frac{4\sqrt{11}}{13}\log s ,
\end{equation}
so that
\begin{equation}
    V_{\rm asym}(\Phi)
    =
    -V_0 e^{-\lambda\Phi},
    \qquad
    \lambda=\frac{8}{\sqrt{11}}.
\end{equation}

Using the general result \eqref{eq:Pabs-scaling} of Section~3, for $d=3$ this gives
\begin{equation}
    P_{\rm abs}
    \sim
    \omega^{16/5},
\end{equation}
which vanishes in the low-energy limit. The effective transverse
dimension is
\begin{equation}
    d_{\rm eff}
    =
    1+\frac{4}{\lambda^2-4}
    =
    \frac{16}{5},
\end{equation}
and hence
\begin{equation}
    \sigma_{\rm abs}
    \sim
    \frac{P_{\rm abs}}{\omega^{d_{\rm eff}-1}}
    \sim
    \omega .
\end{equation}
The absorption cross-section therefore also vanishes as $\omega\to0$,
showing that these vacua exhibit strong decoupling.

It is interesting to contrast this with the examples discussed above.
These vacua do not show integer conformal dimensions, and the full
scalar potential along the flux-backtracking trajectory does not take
the form of a generalised Freund--Rubin potential. Only in the
asymptotic region, where the contribution from the unbounded $F_4$ flux
can be neglected, does the potential reproduce the asymptotic behaviour
of a hypothetical generalised Freund--Rubin compactification with an
effective
\begin{equation}
    \mathrm{AdS}_3\times S^{11/5}
\end{equation}
geometry. Correspondingly, $1+11/5=16/5$ agrees with the effective
transverse dimension found above.
\subsubsection{Decoupling for massive Type IIA on Joyce-type \texorpdfstring{$G_2$}{G22} manifolds}

For these vacua, an explicit brane dual can be constructed, see \cite{ApersAggarwalFarakos2026} for details. The asymptotic steepness along the
flux-backtracking trajectory is
\begin{equation}
    \lambda
    =
    4\sqrt{\frac{2}{5}},
    \qquad
    \lambda^2=\frac{32}{5}.
\end{equation}
Using the general result of Section~3, the corresponding low-frequency
absorption probability behaves as
\begin{equation}
    P_{\rm abs}
    \sim
    \omega^{8/3},
\end{equation}
and therefore vanishes as $\omega\to0$. The effective transverse
dimension is
\begin{equation}
    d_{\rm eff}
    =
    \frac{8}{3},
\end{equation}
so that the absorption cross-section scales as
\begin{equation}
    \sigma_{\rm abs}
    \sim
    \omega .
\end{equation}
These vacua therefore again have strong decoupling in the low-energy
limit.

It is interesting that these vacua also have integer conformal
dimensions. Moreover, the scalar potential along the flux-backtracking
trajectory takes the same form as that of a generalised Freund--Rubin
compactification with an effective
\begin{equation}
    \mathrm{AdS}_3\times S^{5/3}
\end{equation}
geometry. This provides another example in which the integer-dimensional
AdS flux vacua appear to fit naturally into the generalised
Freund--Rubin picture.

\subsection{Guarino--Jafferis--Varela  AdS\texorpdfstring{$_4$}{44} vacua in massive type IIA}

As a further classical example, we consider the AdS$_4$ vacua of
Guarino, Jafferis and Varela (GJV) \cite{GuarinoJafferisVarela2015},
which arise from the consistent truncation of massive type IIA
supergravity on $S^6$. Although these vacua are not scale separated,
they are particularly interesting for our purposes because their
asymptotic geometry shares the asymptotic peculiarity with the DGKT brane
background. As noted in \cite{Apers:2026lgi}, the non-compact
directions parallel to the branes shrink below the string scale in the
far-asymptotic region. They are therefore subject to the same
asymptotic concern emphasized for scale-separated AdS vacua by Bedroya and Steinhardt
\cite{Bedroya:2025ltj}. The GJV solutions nevertheless have a
well-understood holographic interpretation: they are dual to
three-dimensional Chern--Simons--matter theories, for which the
gravitational free energy,
\begin{equation}
    F_{\rm grav}\sim N^{5/3}m^{1/3},
\end{equation}
with $m$ the Romans mass, precisely agrees with the field-theory
partition function \cite{GuarinoJafferisVarela2015}. The corresponding
brane picture consists of $N$ D2-branes deformed by a non-zero Romans
mass. 

An explicit analytic geometry interpolating between the D2-brane region
and the full GJV AdS$_4$ vacuum is not presently available. However, as
in the examples above, the full geometry is not required to determine
the low-energy decoupling properties. Flux backtracking
\cite{Apers:2025pon} determines the asymptotic domain-wall
solution directly. After removing the flux sourced by the D2-branes,
the asymptotic dynamics is controlled by the curvature of $S^6$ and the
Romans mass. Along the resulting flow,
\begin{equation}
    u(r)\sim r^{2/5},
    \qquad
    s(r)\sim r^{4/5},
\end{equation}
the canonically normalised modulus may be chosen as
\begin{equation}
    \Phi=\frac{\sqrt{38}}{5}\log r .
\end{equation}
The asymptotic scalar potential therefore takes the single-exponential
form
\begin{equation}
    V_{\rm asym}(\Phi)
    =
    -V_0e^{-\sqrt{50/19}\,\Phi},
\end{equation}
so that
\begin{equation}
    \lambda^2=\frac{50}{19}.
\end{equation}

Using the general result \eqref{eq:Pabs-scaling} with $d=4$, the absorption probability behaves at low frequency as
\begin{equation}
    \mathcal P_{\rm abs}
    \sim\omega^{25/3}.
\end{equation}
The corresponding effective transverse dimension is
\begin{equation}
    d_{\rm eff}
    =
    1+\frac{4(d-2)}
    {(d-2)\lambda^2-4}
    =
    \frac{22}{3},
\end{equation}
and hence the absorption cross-section scales as
\begin{equation}
    \sigma_{\rm abs}
    \sim
    \frac{\mathcal P_{\rm abs}}
         {\omega^{d_{\rm eff}-1}}
    \sim\omega^2 .
\end{equation}
Thus both the absorption probability and the absorption cross-section
vanish in the low-energy limit, showing that the D2-brane degrees of
freedom decouple from the asymptotic bulk despite the unusual
far-asymptotic behaviour.

\textcolor{black}{Unlike the integer-dimension examples above, the value $d_{\mathrm{eff}}=22/3$ does not have as clean an
interpretation beyond the asymptotic regime. Only asymptotically, there is an effective potential contribution from a $19/3$-dimensional sphere. Near the stabilized moduli, the fractional-sphere picture breaks down.}

\subsection{KKLT vacua}

As a final example, we consider the simplest one-modulus KKLT model. This provides
an illustration of the absorption analysis in a case where the asymptotic
potential is steeper than an ordinary exponential, and therefore approaches the
kination-like regime discussed in Section~3.

We take a single K\"ahler modulus
\begin{equation}
    T=\tau+i\alpha,
    \qquad
    \tau=e^{\sqrt{\frac{2}{3}}\phi},
\end{equation}
where $\phi$ is the canonically normalised volume modulus. The K\"ahler potential
and superpotential are
\begin{equation}
    K=-3\log(2\tau),
    \qquad
    W=W_0+A e^{-aT},
\end{equation}
with $a>0$. The corresponding scalar potential, after fixing the axion at its
minimum, can be written as
\begin{equation}
    V(\tau)
    =
    \frac{a A e^{-a\tau}}{6\tau^2}
    \left[
        A e^{-a\tau}(a\tau+3)+3W_0
    \right].
\end{equation}

At large volume, $\tau\rightarrow\infty$, the leading contribution is therefore
\begin{equation}
    V(\tau)
    \sim
    \frac{a A W_0}{2\tau^2}e^{-a\tau},
\end{equation}
 which is doubly exponential in the canonically normalised
volume modulus.

The corresponding asymptotic domain-wall solution can also be obtained analytically. 
Writing the four-dimensional metric as
\begin{equation}
    ds_4^2 = dr^2 + a(r)^2 ds^2_{\mathbb{R}^{1,2}},
\end{equation}
the asymptotic solution behaves as \cite{Ceresole:2006iq}
\begin{equation}
    a(r) \sim a_0 (r-r_0)^{1/3},
    \qquad
    \phi(r) \sim \frac{1}{\sqrt{3}}
    \log\!\left[3c(r-r_0)\right].
\end{equation}
Upon uplifting this solution, the ten-dimensional metric asymptotically takes the
schematic form
\begin{equation}
    ds_{10}^2
    \sim
    dr^2
    + r^{-2/3} ds^2_{\mathbb{R}^{1,2}}
    + r^{2/3} ds^2_{\rm CY}.
\end{equation}
Note that this is the same metric as the Kasner metric in \cite{ApersConlonMosnyRevello}. Thus, while the internal space grows towards the asymptotic region, the 
length scale along the non-compact directions parallel to the branes decreases as
$r^{-1/3}$. At sufficiently large radial distance this scale becomes smaller than
the string length, which shows that these vacua do not satisfy the decoupling criterion by \cite{Bedroya:2025ltj}.

For $d=4$, the general kination result gives
\begin{equation}
    q=\frac{1}{3},
    \qquad
    d_{\rm eff}=2,
\end{equation}
and the leading low-frequency absorption probability is
\begin{equation}
    \mathcal{P}^{\rm LO}_{\rm abs}
    \sim \omega^3 .
\end{equation}
The corresponding absorption cross-section scales as
\begin{equation}
    \sigma_{\rm abs}
    \sim
    \frac{\mathcal{P}_{\rm abs}}{\omega}
    \sim \omega^2 .
\end{equation}

Hence both the absorption probability and the absorption cross-section vanish in
the low-energy limit, indicating decoupling in this limit.

\newpage
\section{Conclusions}

In this paper, we have studied the low-energy decoupling of branes sourcing AdS
flux vacua. To do so, we considered minimally coupled scalar perturbations and
computed the probability for a wave sourced near the branes to reach the
asymptotic region. In a one-modulus truncation, the result depends on the dimension $d$ of the AdS throat and the steepness $\lambda$
of the asymptotic scalar potential. The different regimes are summarized in
Table~\ref{tab:emission-regimes}.

In all cases the emission probability vanishes in the low-energy limit,
$\omega\rightarrow0$. For sufficiently shallow potentials the suppression is
even stronger: at the critical value of $\lambda$ there are no propagating modes
below a finite threshold, while for still shallower potentials no flux reaches
the asymptotic region at any frequency.

Whenever the asymptotic geometry supports propagating modes,
\begin{equation}
    \lambda>\frac{2}{\sqrt{d-2}},
\end{equation}
we defined an effective transverse dimension
$d_{\rm eff}$ from the asymptotic wave equation. This allows us to define
an absorption cross-section. Although both $\mathcal{P}_{\rm abs}$ and
$d_{\rm eff}$ depend on the steepness of the potential, this dependence cancels
in the cross-section, giving the universal low-energy scaling
\begin{equation}
    \sigma_{\rm abs}\sim\omega^{d-2}.
\end{equation}
This provides successful decoupling for all AdS flux vacua with $d>2$ in the low-energy limit. We show the decoupling properties of different vacua and D-brane systems in Table \ref{tab:examples}.

For Freund–Rubin vacua, \(d_{\mathrm{eff}}\) has a simple geometric meaning: it is the dimension of the internal manifold plus one, or equivalently the dimension of the space explored by the waves. Here we identify a generalised class of Freund-Rubin vacua with a universal modulus of conformal dimension \(\Delta=2(d-1)\). The effective potential for this universal modulus can be obtained by compactifying higher-dimensional gravity on a sphere of possibly fractional dimension, although this construction need not have a geometric interpretation. DGKT and some of the \(\mathrm{AdS}_3\) examples with integer dimensions fall into this class and are shown in bold in Table \ref{tab:examples}.

\begin{table}[ht!]
    \centering
    \small
    \renewcommand{\arraystretch}{1.22}

    \begin{tabular*}{\textwidth}{
        @{\extracolsep{\fill}}
        l c c c c c
        @{}
    }
        \toprule
        Model
        & $\lambda$
        & $d_{\rm eff}$
        & $\mathcal{P}_{\rm abs}^{\rm LO}$
        & $\sigma_{\rm abs}$
        & Dec. \\
        \midrule

        \multicolumn{6}{l}{\emph{Standard brane systems}} \\[1mm]

        D6
        & $\sqrt{\frac{8}{3}}$
        & $3$
        & $\omega^{0}$
        & $\omega^{-2}$
        & No \\

        D5/NS5
        & $\sqrt{\frac{32}{15}}$
        & $4$
        & $\omega^{8}$
        & $\omega^{5}$
        & Strong \\

        D4
        & $\sqrt{2}$
        & $5$
        & $\omega^{8}$
        & $\omega^{4}$
        & Strong \\

        \textbf{D3 / AdS$_5\times S^5$}
        & $\sqrt{\frac{32}{15}}$
        & $6$
        & $\omega^{8}$
        & $\omega^{3}$
        & Strong \\

        D2
        & $\sqrt{\frac{8}{3}}$
        & $7$
        & $\omega^{8}$
        & $\omega^{2}$
        & Strong \\

        D1
        & $\sqrt{\frac{32}{7}}$
        & $8$
        & $\omega^{8}$
        & $\omega$
        & Strong \\

        \textbf{M2 / AdS$_4\times S^7$}
        & $\sqrt{\frac{18}{7}}$
        & $8$
        & $\omega^{9}$
        & $\omega^{2}$
        & Strong \\

        \textbf{M5 / AdS$_7\times S^4$}
        & $\sqrt{\frac{3}{5}}$
        & $5$
        & $\omega^{9}$
        & $\omega^{5}$
        & Strong \\

        M2--M2--M5--M5 / AdS$_2$
        & --
        & --
        & $\omega^{2}$
        & $\omega^{0}$
        & Weak \\

        \midrule
        \multicolumn{6}{l}{\emph{AdS flux vacua}} \\[1mm]

        \textbf{DGKT--CFI AdS$_4$}
        & $\sqrt{\frac{54}{13}}$
        & $\frac{20}{7}$
        & $\omega^{27/7}$
        & $\omega^{2}$
        & Strong \\

        \textbf{SS AdS$_4$ in IIA}
        & $\sqrt{\frac{54}{13}}$
        & $\frac{20}{7}$
        & $\omega^{27/7}$
        & $\omega^{2}$
        & Strong \\

        \textbf{SS AdS$_3$ in IIB ($n_2$)}
        & $4\sqrt{\frac{3}{7}}$
        & $\frac{12}{5}$
        & $\omega^{12/5}$
        & $\omega$
        & Strong \\

        SS AdS$_3$ in mIIA (singular $G_2$)
        & $\frac{8}{\sqrt{11}}$
        & $\frac{16}{5}$
        & $\omega^{16/5}$
        & $\omega$
        & Strong \\

        \textbf{SS AdS$_3$ in mIIA (Joyce $G_2$)}
        & $4\sqrt{\frac{2}{5}}$
        & $\frac{8}{3}$
        & $\omega^{8/3}$
        & $\omega$
        & Strong \\

        GJV AdS$_4$
        & $\sqrt{\frac{50}{19}}$
        & $\frac{22}{3}$
        & $\omega^{25/3}$
        & $\omega^{2}$
        & Strong \\

        KKLT
        & $\lambda_{\rm eff}\to\infty$
        & $2$
        & $\omega^{3}$
        & $\omega^{2}$
        & Strong \\

        \bottomrule
    \end{tabular*}

    \caption{
    Leading low-frequency absorption probability, effective transverse dimension,
    absorption cross-section and decoupling behaviour. ``SS'' denotes scale
    separated. Weak decoupling corresponds to
    $\mathcal{P}_{\rm abs}\to0$, while strong decoupling additionally requires
    $\sigma_{\rm abs}\to0$. Models admitting an ordinary or generalised
    Freund--Rubin interpretation are shown in bold.
    }
    \label{tab:examples}
\end{table}

\subsection*{Future directions}

There are several interesting questions for future work. 
\begin{itemize}
    \item \textcolor{black}{
Our analysis points to several different types of decoupling. We can distinguish between weak and strong decoupling, depending on whether only the absorption probability or also the absorption cross-section vanishes. There is also a difference in the range of energies over which decoupling occurs: in some cases it is strictly a low-energy phenomenon, while in others it persists over a finite range of energies, or even at all frequencies. Any of these behaviours should in principle be sufficient to isolate a decoupled CFT, but it would be interesting to understand whether the precise type and rate of decoupling are reflected in the dual theory.
In the regime where asymptotic scattering states exist, we find
\[
P_{\rm abs} \sim \omega^{\Delta_{\rm abs}}, \qquad \Delta_{\rm abs} = d+d_{\rm eff}-3.
\]
It is natural to ask whether the exponent $\Delta_{\rm abs}$ has a direct interpretation in the CFT. This is particularly suggestive for DGKT and the generalised Freund--Rubin examples, where $\Delta_{\rm abs}$ is related to the effective dimension and this structure appears to characterize the full one-modulus trajectory, rather than only its asymptotic limit. It would be interesting to understand whether this gives $\Delta_{\rm abs}$ a more concrete meaning on the CFT side.}
\item \textcolor{black}{Another direction is to explore more systematically the relation between these brane geometries and FLRW cosmologies. As emphasized in \cite{Bedroya:2025ltj}, the asymptotic radial solutions map to scalar-driven scaling cosmologies. Since the exponential potentials relevant for scale-separated backgrounds are rather steep, it would be interesting to ask whether these asymptotics remain stable once radiation-like perturbations are included, in analogy with the kination and tracker dynamics discussed in \cite{ConlonRevello, ApersConlonMosnyRevello,
ApersConlonMosnyUplifts,ApersConlonCopelandMosnyRevello,
MosnyConlonCopeland}. A small radiation component can grow relative to the scalar background and eventually drive the system towards a radiation tracker. From the brane perspective, one may then ask whether an outgoing perturbation can backreact strongly enough that the far asymptotic geometry is no longer described by the pure scalar scaling solution.}

\item Although we have established low-energy decoupling, the brane geometries still exhibit the unusual asymptotic behaviour emphasized in \cite{Bedroya:2025ltj}: sufficiently far from the branes, perturbations eventually probe energies above the string scale. It would be useful to understand whether this has any real consequence for the holographic dual. The Guarino--Jafferis--Varela \cite{GuarinoJafferisVarela2015} vacua provide a good testing ground, since they appear to share this asymptotic behaviour while having a well-understood CFT dual. \textcolor{black}{Interestingly, there may also exist a class of scale-separated AdS$_3$ vacua in massive type IIA for which the brane dual is realized on a finite radial interval, so that no asymptotic region at infinite distance is ever reached. Such constructions could therefore avoid the issues associated with the far-asymptotic regime altogether \cite{ApersAggarwalFarakos2026}.}

\item \textcolor{black}{The one-modulus trajectories considered here arise naturally for cohomogeneity-one domain walls, where all scalars depend on a single radial coordinate. It would be interesting to study genuinely multi-moduli brane configurations, for instance along the lines of the intersecting end-of-the-world branes of \cite{Angius:2023xtu}, and to understand how such configurations decouple from the asymptotic bulk.}
\item The brane descriptions now available for DGKT and several scale-separated AdS$_3$ vacua provide a starting point for extracting more detailed properties of their putative dual CFTs.
\end{itemize}

\section*{Acknowledgements}

We are grateful to Alek Bedroya, Miguel Montero,
Muthusamy Rajaguru, Irene Valenzuela, and Vincent Van Hemelryck for useful conversations on this topic. We also thank Joseph Conlon and Edward Hardy for valuable comments on the draft. 

F.A. is supported by the ERC Starting Grant QGuide101042568 -- StG 2021.
N.S.G. acknowledges support from the Oxford-Berman Graduate Scholarship jointly
funded by the Clarendon Fund and the Rudolf Peierls Centre for Theoretical Physics
Studentship.

\appendix

\section{Generalised Freund--Rubin vacua}
\label{app:generalised-freund-rubin}

We consider a $(d+n)$-dimensional theory of gravity coupled to an
$n$-form field strength,
\begin{equation}
    S_{d+n}
    =
    \frac{1}{2\kappa_{d+n}^2}
    \int d^{d+n}x\,\sqrt{-g_{d+n}}
    \left(
        R_{d+n}
        -
        \frac{1}{2\,n!}F_n^2
    \right).
    \label{eq:FR-higher-dimensional-action}
\end{equation}
We compactify the theory on an $n$-sphere threaded by $N$ units of
$n$-form flux,
\begin{equation}
    \int_{S^n} F_n \propto N ,
\end{equation}
where numerical factors associated with flux quantization are absorbed
into the definition of $N$. We are interested in the dynamics of the
overall internal volume, or breathing mode.

We therefore write the $(d+n)$-dimensional metric as
\begin{equation}
    ds_{d+n}^2
    =
    \mathcal{V}^{-\frac{2}{d-2}} ds_d^2
    +
    \mathcal{V}^{\frac{2}{n}} d\widetilde{s}_n^2 ,
    \label{eq:FR-metric-ansatz}
\end{equation}
where $d\widetilde{s}_n^2$ is a fixed reference metric on the internal
sphere with unit volume, and $\mathcal{V}$ denotes its physical volume
in these units. The prefactor multiplying $ds_d^2$ has been chosen such
that the lower-dimensional metric is in $d$-dimensional Einstein frame.

Neglecting derivatives of $\mathcal{V}$ for the moment, the
higher-dimensional Ricci scalar decomposes as
\begin{equation}
    R_{d+n}
    =
    \mathcal{V}^{\frac{2}{d-2}} R_d
    +
    \mathcal{V}^{-\frac{2}{n}} \widetilde{R}_n ,
    \label{eq:FR-Ricci-split}
\end{equation}
where $\widetilde{R}_n>0$ is the Ricci scalar of the reference sphere.
For flux threading the internal space we may write
\begin{equation}
    F_n = N\,\widetilde{\mathrm{vol}}_n ,
\end{equation}
so that
\begin{equation}
    F_n^2
    =
    n!\,N^2\,\mathcal{V}^{-2}.
    \label{eq:FR-flux-scaling}
\end{equation}
The $d$-dimensional effective potential for the breathing mode is
therefore
\begin{equation}
    V(\mathcal{V})
    =
    -\widetilde{R}_n\,
    \mathcal{V}^{-\frac{2(n+d-2)}{n(d-2)}}
    +
    \frac{1}{2}N^2\,
    \mathcal{V}^{-\frac{2(d-1)}{d-2}} .
    \label{eq:FR-volume-potential}
\end{equation}
The first term arises from the positive curvature of the internal
sphere, while the second is the positive energy density carried by the
$n$-form flux.

Including derivatives of the volume modulus, dimensional reduction
gives the kinetic term
\begin{equation}
    \mathcal{L}_{\rm kin}
    =
    -
    \frac{d+n-2}{n(d-2)}
    \frac{1}{\mathcal{V}^2}
    \partial_\mu\mathcal{V}\,
    \partial^\mu\mathcal{V}.
    \label{eq:FR-volume-kinetic}
\end{equation}
Thus, in the normalisation used here, the canonically normalised
breathing mode is
\begin{equation}
    \varphi
    =
    \sqrt{\frac{d+n-2}{n(d-2)}}
    \ln \mathcal{V} .
    \label{eq:FR-canonical-modulus}
\end{equation}
Equivalently,
\begin{equation}
    \mathcal{V}
    =
    \exp\left[
        \sqrt{\frac{n(d-2)}{d+n-2}}\,
        \varphi
    \right].
\end{equation}
In terms of the canonically normalised field $\varphi$, the scalar
potential can therefore be written explicitly as
\begin{equation}
    \boxed{
    V(\varphi)
    =
    -\widetilde{R}_n\,
    \exp\left[
        -2\sqrt{\frac{d+n-2}{n(d-2)}}\,\varphi
    \right]
    +
    \frac{1}{2}N^2\,
    \exp\left[
        -2(d-1)
        \sqrt{\frac{n}{(d-2)(d+n-2)}}\,\varphi
    \right]
    } .
    \label{eq:FR-canonical-potential}
\end{equation}

Extremizing Eq.~\eqref{eq:FR-volume-potential} gives
\begin{equation}
    \mathcal{V}_\star
    =
    \left[
        \frac{n(d-1)N^2}
        {2(n+d-2)\widetilde{R}_n}
    \right]^{\frac{n}{2(n-1)}} .
    \label{eq:FR-stabilized-volume}
\end{equation}
At large flux, the stabilized internal volume therefore scales as
\begin{equation}
    \mathcal{V}_\star
    \sim
    N^{\frac{n}{n-1}} .
\end{equation}

The vacuum energy scales as
\begin{equation}
    V_\star
    \sim
    -
    N^{-\frac{2(n+d-2)}{(n-1)(d-2)}} ,
\end{equation}
so the Freund--Rubin extremum is an AdS vacuum.

At the extremum one furthermore finds
\begin{equation}
    \left.
    \frac{V''(\varphi)}{V(\varphi)}
    \right|_{\varphi_\star}
    =
    -
    4\frac{d-1}{d-2},
    \label{eq:FR-second-derivative}
\end{equation}
such that the mass with respect to the AdS radius is given by
\begin{equation}
    m_\varphi^2 L_{\rm AdS}^2
    =
    2(d-1)^2 ,
    \label{eq:FR-breathing-mode-mass}
\end{equation}
which is independent of the flux or the dimensionality of the sphere.
The dual conformal dimension is then
\begin{equation}
    \boxed{
    \Delta = 2(d-1)
    } .
    \label{eq:FR-breathing-mode-dimension}
\end{equation}
Note that this is the same integer conformal dimension found in \cite{Apers:2022zjx2} for AdS flux vacua with a simple large-flux brane realization. This suggests that such vacua may be viewed, at least in an effective sense, as generalised Freund--Rubin vacua. As also noted in \cite{Apers:2022zjx2}, the appearance of this integer dimension implies, in the large-$N$ limit, the presence of an emergent level-$k=d-1$ polynomial shift symmetry for the breathing mode for these generalised Freund-Rubin vacua.

\section{Details on the absorption computation for AdS\texorpdfstring{$_3$}{33}}\label{app:ads3-absorption}
\subsection{The flux vacua}

We consider the supersymmetric scale-separated AdS$_3$ vacua of
\cite{VanHemelryck:2025qok} obtained from type IIB string theory on the
nilmanifold
\begin{equation}
    \mathfrak n_2 \simeq {\rm Nil}_3 \times \mathbb T^4 .
\end{equation}
The compactification preserves minimal $\mathcal N=1$ supersymmetry in three
dimensions. The ingredients are RR three-form flux $F_3$, RR
seven-form flux $F_7$, and the metric flux $\omega$ associated with the
twisting of the nilmanifold. In the conventions used here, the independent
flux components are denoted by $F_{31},\ldots,F_{37}$ and $F_{71}$.

The $F_3$ components $F_{32}$ and $F_{33}$ are constrained by the O5-plane tadpole. By contrast,
\begin{equation}
    F_{71},
    \qquad
    F_{31},\,F_{34},\,F_{35},\,F_{36},\,F_{37},
\end{equation}
are independently unbounded. Parametric scale separation is generated by
taking the unbounded three-form fluxes large, while $F_{71}$ is scaled
appropriately in order to simultaneously obtain large internal radii and
weak string coupling.

 The scalar sector is parametrized by the three-dimensional dilaton $s$ and
the seven cycle sizes $L_i$, $i=1,\ldots,7$, with
\begin{equation}
    s^{-1}
    =
    g_s^{-1}\sqrt{\mathcal V_7},
    \qquad
    \mathcal V_7\equiv \prod_{i=1}^7 L_i .
\end{equation}
The real superpotential takes the form
\begin{equation}
    P
    =
    -\frac{s^3}{4\sqrt{\mathcal V_7}}\,\mathcal F
    +\frac{s^2\omega L_7}{4L_1L_2},
    \label{eq:n2-superpotential-app}
\end{equation}
where
\begin{align}
    \mathcal F={}&
    F_{71}
    -F_{33}L_1L_2L_3L_4
    -F_{32}L_1L_2L_5L_6
    +F_{31}L_3L_4L_5L_6
    \nonumber\\
    &+F_{37}L_1L_3L_5L_7
    +F_{36}L_2L_4L_5L_7
    +F_{35}L_2L_3L_6L_7
    +F_{34}L_1L_4L_6L_7 .
    \label{eq:n2-flux-combination-app}
\end{align}
The scalar-field metric is
\begin{equation}
    G_{AB}
    =
    {\rm diag}\left(
    \frac{2}{s^2},
    \frac{1}{2L_1^2},\ldots,\frac{1}{2L_7^2}
    \right),
\end{equation}
such that
\begin{equation}
    \mathcal L_{\rm kin}
    =
    -\frac{2}{s^2}(\partial s)^2
    -\frac12\sum_{i=1}^{7}\frac{(\partial L_i)^2}{L_i^2}.
\end{equation}
The scalar potential is obtained from the real superpotential as
\begin{equation}
    V=G^{AB}\partial_A P\,\partial_B P-4P^2 .
\end{equation}

At the supersymmetric minimum, the eight scalar masses are independent of
the fluxes and are given in AdS units by
\begin{equation}
    m^2L_{\rm AdS}^2
    =
    \left\{
        120,\,
        8,8,8,8,8,8,8
    \right\}.
    \label{eq:n2-mass-spectrum}
\end{equation}
The corresponding conformal dimensions are
\begin{equation}
    \Delta
    =
    \left\{
        12,\,
        4,4,4,4,4,4,4
    \right\}.
    \label{eq:n2-conformal-dimensions}
\end{equation}
\subsection{Background metric}

We now apply the flux-backtracking prescription
\cite{Apers:2025pon} to the type IIB AdS$_3$ vacua on
$\mathfrak n_2$. The nilmanifold is defined by
\begin{equation}
    de^7=\omega\,e^{12},
    \qquad
    de^i=0,
    \quad i=1,\ldots,6.
\end{equation}

Upon removing the unbounded fluxes, the AdS$_3$ critical point is replaced
by a running domain-wall solution. The corresponding BPS flow is
\begin{equation}
    e^\phi=r^{-1/3},
    \qquad
    L_1=L_2=r^{1/12},
    \qquad
    L_3=L_4=L_5=L_6=1,
    \qquad
    L_7=r^{-1/12},
\end{equation}
together with
\begin{equation}
    A(r)=\frac{7}{12}\log r .
\end{equation}
Introducing $\rho=r^{1/4}$, the ten-dimensional string-frame geometry
takes the form
\begin{align}
    ds^2_{10,S}
    ={}&d\rho^2+\rho^{-4/3}dx_\mu dx^\mu
    +\rho^{2/3}\big[(e^1)^2+(e^2)^2\big]
    \nonumber\\
    &+\sum_{i=3}^{6}(e^i)^2
    +\rho^{-2/3}(e^7)^2,
    \label{eq:n2-backtracked-metric}
    \\
    e^\phi={}&\rho^{-4/3}.
    \label{eq:n2-backtracked-dilaton}
\end{align}
The backtracked geometry ends at the singular locus $\rho=0$, where the
string coupling diverges. At this endpoint the $1,2$ directions shrink,
the $3,\ldots,6$ directions remain finite, and the Heisenberg fibre $e^7$
expands. This is the locus at which the branes dual to the removed fluxes
are placed.

We then restore the unbounded fluxes by introducing the corresponding
D1- and D5-branes, whose configuration is summarized in
Table~\ref{tab:D1D5system}. Denoting the D1 harmonic function by $H$ and,
for the symmetric branch, the common D5 harmonic function by $K$, the
resulting interpolating geometry is
\begin{align}
    ds^2_{10,S}
    ={}&(HK^5)^{1/2}d\rho^2
    +(HK^5)^{-1/2}\rho^{-4/3}dx_\mu dx^\mu
    +(HK)^{1/2}\rho^{2/3}
      \big[(e^1)^2+(e^2)^2\big]
    \nonumber\\
    &+(HK^{-1})^{1/2}
      \sum_{i=3}^{6}(e^i)^2
    +(HK^{-3})^{1/2}\rho^{-2/3}(e^7)^2,
    \label{eq:n2-D1D5-metric}
    \\
    e^\phi={}&(HK^{-5})^{1/2}\rho^{-4/3},
     \label{VHmetric}
\end{align}
with
\begin{equation}
    H(\rho)=1+\frac{N_1}{\rho^{2/3}},
    \qquad
    K(\rho)=1+\frac{N_5}{\rho^{2/3}} .
\end{equation}
This geometry interpolates between the flux-backtracked solution at large
$\rho$ and the original AdS$_3$ vacuum in the near-horizon region.
\begin{table}[t]
\centering
\begin{tabular}{c|c c c|c c c c c c c}
 & $x_1$ & $x_2$ & $r$ &
 $e_1$ & $e_2$ & $e_3$ & $e_4$ & $e_5$ & $e_6$ & $e_7$ \\
\hline
D1    & $\otimes$ & $\otimes$ & - & - & - & - & - & - & - & - \\
\hline
D5$_1$ & $\otimes$ & $\otimes$ & - & - & - & $\otimes$ & $\otimes$ & $\otimes$ & $\otimes$ & - \\
\hline
D5$_4$ & $\otimes$ & $\otimes$ & - & $\otimes$ &- & - &  $\otimes$&-& $\otimes$ &  $\otimes$ \\
\hline
D5$_5$ & $\otimes$ & $\otimes$ & - & - & $\otimes$ & $\otimes$ & - & - & $\otimes$ & $\otimes$ \\
\hline
D5$_6$ & $\otimes$ & $\otimes$ & - & - & $\otimes$ & - & $\otimes$ & $\otimes$ & - & $\otimes$ \\
\hline
D5$_7$ & $\otimes$ & $\otimes$ & - & $\otimes$ & - & $\otimes$ & - & $\otimes$ & - & $\otimes$ \\
\end{tabular}
\caption{D-brane system associated to the scale-separated AdS$_3$ vacua obtained from compactifying IIB string theory on a nilmanifold $\mathfrak{n}_2$ \cite{VanHemelryck:2025qok}}
\label{tab:D1D5system}
\end{table}
\subsection{Wave equation}
We will analyse the scalar wave equation in the 3-dimensional Einstein frame. Given the 10-dimensional metric in the string frame in Eq. \eqref{eq:n2-D1D5-metric}, the corresponding metric in the 3-dimensional Einstein frame is
\begin{align}
    g^{(3,E)}_{MN}= \left(\mathcal{V}_{7}^{(s)}\right)^{2}\cdot e^{-4\phi}\cdot  g^{(10,s)}_{MN} = (HK^5)^{3/2} \cdot \rho^6\cdot  g^{(10,s)}_{MN}.
\end{align}
More explicitly, 
\begin{align}
    ds_{(10,E)}^2
    ={}&(HK^5)^2 \rho^6\, d\rho^2
    +(HK^5)\rho^{14/3}\, dx_\mu dx^\mu
    +(HK^4)^2\rho^6\big[(e^1)^2+(e^2)^2\big]
    \nonumber\\
    &+H^2K^7\rho^6
    \big[(e^3)^2+(e^4)^2+(e^5)^2+(e^6)^2\big]
    +(HK^3)^2\rho^{10/3}(e^7)^2 .
\end{align}
In a general spacetime given by a metric $g_{MN}$, the equation of motion for a scalar s-wave $\Phi(t,\rho) = \phi(\rho) e^{i\omega t}$ is 
\begin{align}
    0&= \phi''(\rho) + \frac{g_{\rho\rho}}{\sqrt{-g}}\partial_\rho \left(\frac{\sqrt{-g}}{g_{\rho\rho}}\right)\cdot  \phi'(\rho) + \left(\frac{g_{\rho\rho}}{-g_{tt}}\right)\cdot\omega^2  \phi(\rho).
\end{align}
For the D1-D5 system under consideration, this takes the form below 
\begin{align}
    0&= \phi''(\rho) + \frac{5}{3\rho}  \phi'(\rho) + \omega^2 \; (HK^5) \rho^{4/3}\; \phi(\rho).
\end{align}
Defining $\phi(\rho) = \rho^{-5/6} \psi(\rho)$, the wave equation takes the form
\begin{align}\label{scalarwaveeqn}
    0&= \psi''(\rho) +\left[\frac{5}{36\rho^{2}}+ \omega ^2(HK^5) \rho^{4/3}  \right]\psi(\rho).
\end{align}
We can therefore read off the effective potential
\begin{align}
    V_{\rm eff}(\rho) = -\frac{5}{36\rho^{2}}- \omega ^2  \rho^{4/3} \left[1+\left(\frac{R_{D1}}{\rho}\right)^{2/3}\right]\left[1+\left(\frac{R_{D5}}{\rho}\right)^{2/3}\right]^5.
\end{align}
This potential is shown in Figure~\ref{fig:3dpotbar}.
\subsection{Absorption computation}

\subsubsection{Near-region}
In the near-region limit $\rho\rightarrow 0$, where $H/R_{D1}^{2/3} = K/R_{D5}^{2/3} = \rho^{-2/3}$, the equation becomes
\begin{align}
    0&= \phi''(\rho) + \frac{5}{3\rho}  \phi'(\rho) + \omega^2 L^4\; \rho^{-8/3}\; \phi(\rho),
\end{align}
where we have defined $L\equiv (R_{D1}R_{D5}^5)^{1/6}$. In terms of the AdS proper radial coordinate,
\begin{align}
    z\equiv \frac{3L^2}{\rho^{1/3}},
\end{align}
the 3-dimensional metric takes the form 
\begin{align}
    ds^2_3 = \frac{L^2_{\rm AdS}}{z^2} (-dt^2 + dx^2 +dz^2) ,
\end{align}
where $L_{\rm AdS} \equiv 3L^4$. On the other hand, the equation takes the form of a Bessel differential equation
\begin{align}
    0=z^2 \phi''(z) - z \phi'(z) + (\omega z)^2 \phi(z).
\end{align}
Its general solution for ingoing boundary conditions can be written as 
\begin{align}\label{eq:nearsolD1D5}
    \phi_{\rm near}(z) = c_1\cdot z \cdot H^{(1)}_1 (\omega z ) .
\end{align}
In the limit $\omega z \ll1 $, or equivalently $\omega L^2 / \rho^{1/3} \ll 1$, the solution becomes dominated by the zero mode. Using the small argument expansion for the Hankel function the solution looks as follows, 
\begin{align}
    \phi_{\rm near}(z) = \frac{2c_1}{i\pi\omega} + \mathcal{O}(z^2).
\end{align}
\subsubsection{Far-region}
In the far-region limit $\rho\rightarrow \infty$, where $H= K=1$, the equation becomes
\begin{align}
    0&= \phi''(\rho) + \frac{5}{3\rho}  \phi'(\rho) + \omega^2 \rho^{4/3}\; \phi(\rho).
\end{align}
In this region, the metric takes a domain-wall form
\begin{align}
    ds^2_3 = dR^2 + e^{2A(R)} \cdot \eta_{\mu\nu} d\tilde{x}^\mu d\tilde{x}^\nu,
\end{align}
where $A(R)= q\;\log R$ and $q= 7/12$, and with the rescaling $\tilde{x}^\mu = 4^{7/12} x^\mu$. The proper radial coordinate $R$ in terms of the previous coordinate $\rho$ is 
\begin{align}
    R\equiv \frac{\rho^4}{4}.
\end{align}
In terms of this new coordinate, the equation takes the form
\begin{align}
    0&= \phi''(R) + \frac{7}{6R}  \phi'(R) + \omega^2_\infty R^{-7/6}\; \phi(R),
\end{align}
where $\omega_\infty = 4^{-7/12} \omega $ is defined due to the rescaling of the coordinate $\tilde{t}=4^{7/12} t$, such that $\omega \cdot t = \omega_\infty \cdot \tilde{t}$. Defining the parameters 
\begin{align}
    b= \frac{5}{12}, \hspace{1 em} \xi = \frac{12}{5}\omega_\infty \, R^{5/12}, \hspace{1 em} n= \frac{7}{5}, \hspace{1 em} \nu = \frac{1}{5},
\end{align}
the equation becomes, 
\begin{align}
    0&= \phi''(\xi) + \frac{7}{5\xi}\phi'(\xi)+ \phi(\xi),
\end{align}
with general solution
\begin{align}
    \phi_{\rm far} (\xi) = \xi^{-1/5}\left[c_3 \cdot J_{1/5}(\xi) + c_4\cdot Y_{1/5}(\xi) \right].
\end{align}
In the limit $\omega R^{5/12} \ll 1$, or equivalently $\omega \rho^{5/3} \ll 1$ , the solution becomes dominated by the zero mode. Using the small argument expansion for the Bessel functions, one finds
\begin{align}
     \xi^{-1/5} \cdot J_{1/5}(\xi)&= \frac{2^{-1/5}}{\Gamma(6/5)}   +\mathcal{O}(\xi^{6/5}),
     \\
      \xi^{-1/5}\cdot Y_{1/5}(\xi)&= -\frac{2^{1/5}\Gamma(1/5)}{\xi^{2/5}}- \frac{(1+\sqrt{5})\Gamma(-1/5)}{2^{1/5} 4\pi}+\mathcal{O}(\xi^{6/5}).
\end{align}
\subsubsection{Overlap region}
There is a low-frequency regime in which both solutions reduce to the universal zero-frequency mode. This happens in the region where $\omega L^2 \ll \rho^{1/3}$ and $ \omega \rho^{5/3} \ll 1$ are both true, that is 
\begin{align}
   (\omega L^2)^3 \ll \rho \ll \omega^{-3/5}.
\end{align}
This overlap region exists in a low-frequency regime such that
\begin{align}
 \omega L^{5/3}< 1.
\end{align}
Imposing matching conditions for the far and near solutions in this region gives
\begin{align}
    c_3= \frac{c_1}{i\pi } \frac{2^{6/5} \Gamma(6/5)}{\omega}, \hspace{2 em} c_4 = 0.
\end{align}
As a result, the incoming part of the far solution is 
\begin{align}\label{eq:farsolD1D5}
    \phi_{\rm far}^{\rm in} (\xi) = \frac{c_1}{i\pi } \frac{2^{6/5} \Gamma(6/5)}{\omega} \xi^{-1/5} \cdot \frac{1}{2}\left[J_{1/5}(\xi) + i\;Y_{1/5}(\xi) \right].
\end{align}
\subsubsection{Absorption Probability}
Given the solution in Eq.\eqref{eq:nearsolD1D5} for the scalar wave equation in the near region, the corresponding absorbing flux into the throat is 
\begin{align}
    \mathcal{F}^{\rm in}_{\rm near}=\frac{1}{2i} \sqrt{-g}\,g^{zz}(\phi^*_{\rm near} \partial_z \phi_{\rm near}  -\phi_{\rm near}  \partial_z\phi^*_{\rm near}  )  = \frac{6L^4}{\pi}|c_1|^2.
\end{align}
Similarly, for the solution in Eq.\eqref{eq:farsolD1D5} for the scalar wave equation in the far region, the corresponding incoming flux at infinity
\begin{align}
    \mathcal{F}^{\rm in}_{\rm far}&=\frac{1}{2i} \sqrt{-g}\,g^{RR}(\phi^*_{\rm far} \partial_R \phi_{\rm far}  -\phi_{\rm far}  \partial_R \phi^*_{\rm far}  ) =\frac{\Gamma^2(6/5) }{\pi^3 \omega^{12/5}} \left(\frac{10}{3}\right)^{7/5} |c_1|^2.
\end{align}
As a result, the absorption probability is 
\begin{align}
\mathcal{P}_{\rm abs}\equiv \frac{\mathcal{F}^{\rm in}_{\rm near}}{\mathcal{F}^{\rm in}_{\rm far}}=\frac{9 \left(\frac{3}{10}\right)^{2/5} \pi^2}{5\Gamma^2 \left(\frac{6}{5}\right)}(\omega L^{5/3})^{12/5}.
\end{align}
The general expectations for the absorption fluxes found in Eq. \eqref{eq:ads-absorbing-flux} and Eq.\eqref{eq:flux-far-qless1}, and for the absorption probability in Eq. \eqref{eq:Pabs-unified-alpha-nu}, match the expressions above when evaluated for the parameters describing this system, i.e. $\alpha = 1, \,\nu = 1/5,\,L_{\rm AdS}= 3L^4, c_{\infty}= 4^{-7/12}$.
\bibliographystyle{utphys}
\bibliography{Refs}

\end{document}